\documentclass[12pt]{style_thesis}
\pdfoutput=1
\usepackage{tikz}
\usepackage{graphicx}
\usepackage{mathtools}
\usepackage{lipsum}
\usepackage[utf8]{inputenc}
\usepackage[T1]{fontenc}
\usepackage{newtxtext,newtxmath}
\usepackage[
    backend=biber,natbib,
    style=  numeric, 
    citestyle= numeric 
]{biblatex}
\begin{document}
\title{Development of Deep Learning Methods for Inflow Turbulence Generation}
\author{PATIL Aakash Vijay}
\degreeaward{Master of Science}            
\university{Ecole Centrale de Lille}       
\address{Lille, France}                        
\universitysec{Ecole Nationale Superieure de Mecanique et d'Aerotechnique}        
\addresssec{Poitiers, France}                        
\unilogo{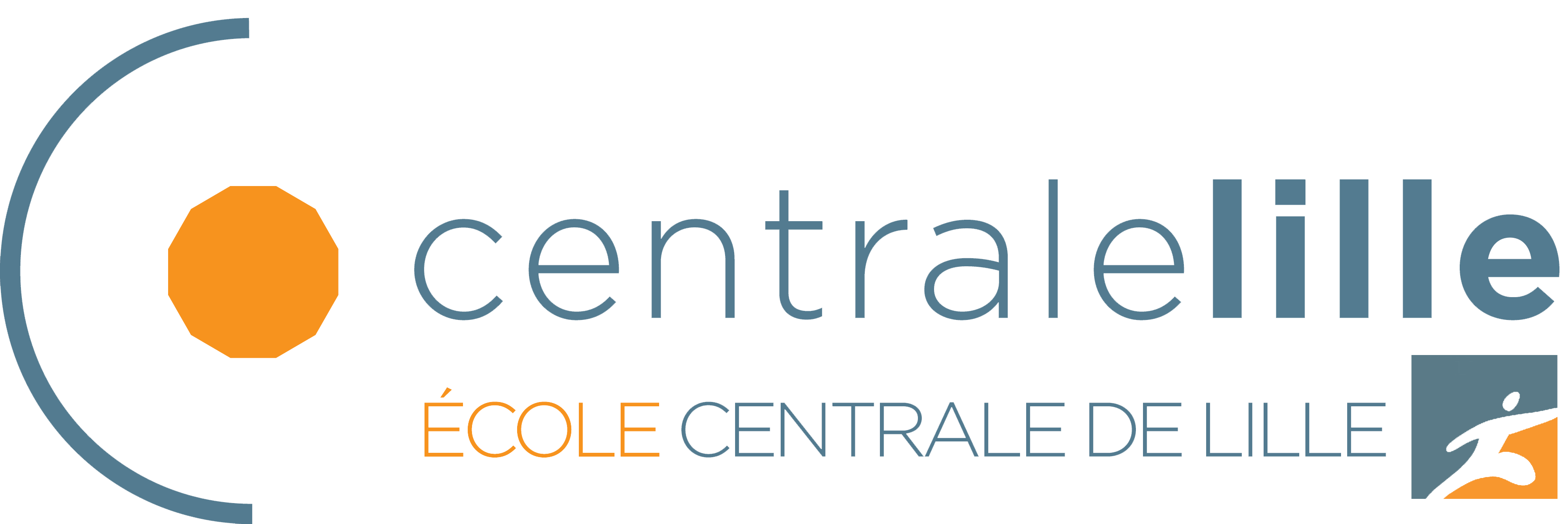}  
\preparedat{CERFACS}  
\preparedataddr{Toulouse, France}        
\copyyear{September 2019}              
\rightsstatement{All rights reserved except where otherwise noted.}
\maketitle[logo]

\newpage
\begin{center}
\thispagestyle{empty}
\vspace*{\fill}
\textit{Dedicated to my best friend,  Dr. Jayesh Zadokar.}
\vspace*{\fill}
\end{center}

\begin{acknowledgements}
I am extremely thankful to Dr. Corentin Lapyere at CERFACS for his supervision during the master's thesis. I am thankful for providing this opportunity to work on a very interesting yet challenging problem, teaching me to think critically, and for answering every small query. I am also thankful to Dr. Guillaume Daviller for his suggestions and guidance throughout the work of this thesis, and to Mr. Blanchard for his kind help. I would especially like to thank CERFACS for providing all the computational resources throughout the work of this thesis. I would also like to thank the members of HELIOS group at CERFACS for their suggestions from time to time, and my colleagues Bastien and Pierre who made the stay memorable. At last, I would like to thank the amazing administrative staff as CERFACS as well as ENSMA who kindly helped to make my stay pleasant in Toulouse.

\end{acknowledgements}

\tableofcontents
\listoffigures
\printnomenclature
\mainmatter

\chapter{Introduction}

\section{Inflow Turbulence }
Generating realistic inflow of turbulence is very challenging. Various possibilities to generate such realistic inflow conditions have been proposed in the literature. Though computationally expensive, the strategy of resolving the transition to turbulence starting from the laminar flow is the only perfect solution for generating exact turbulent inflows \cite{rai1993direct, bardino1983improved}. Another widely used strategy reported in the literature is that of using precursor simulation to generate inflow turbulence which provides an accurate solution to this problem but is also expensive in computation as well large storage is required for storing all temporal data \cite{lund1998generation}. For turbulent boundary layer inflows, a commonly used method is that of recycling turbulent velocity components from the downstream planes and injecting back to the upstream plane of the same domain. Though this method is relatively cheap in terms of computations and memory requirements, obtaining the realistic inflows is challenging as the scalings required for recycling are unknown \cite{bardino1983improved, lund1998generation, araya2011dynamic}. Other methods for generating inflow turbulence, though easy to perform require longer domains to recover correct statistics, include the imposition of random noise on mean velocity profile as well as with vortices from developed turbulence\cite{araya2011dynamic, adams2001deconvolution}. Inflow conditions based on digital filtering techniques, proper orthogonal decomposition, and linear stochastic estimation have also been proposed \cite{keating2004priori, klein2003digital, perret2008turbulent}. The reader may refer to the work of \cite{wu2017inflow} for a detailed review of the methods of generating inflow turbulence. At CERFACS, the AVBP solver uses Kraichnan and Celiek's method of generating patches of turbulence inflow. In this method, the inflow is created with required statistical properties and then added to the mean flow. Kraichnan's method is to generate turbulence inflow with statistical properties equivalent to that of incompressible and isotropic turbulent velocity fields, whereas Celiek's method is an extension with spatially inhomogeneous and anisotropic properties \cite{kraichnan1970diffusion} \cite{smirnov2001random}.

Use of machine learning methods in the area of fluid dynamics is being reported since the works of Duraisamy and Xiao\cite{PARISH2016758}\cite{xiao2016quantifying}, who have demonstrated the use of these methods for turbulence modeling in the form of estimation of model uncertainties using machine learning. There have been some studies where direct prediction of Reynolds stresses for RANS and prediction of deconvoluted direct numerical simulation have been proposed \cite{ling2016reynolds} \cite{maulik2017neural}. Recently, Fukami and Mohan have demonstrated the use of re-generating turbulence statistics from DNS data \cite{fukami2019synthetic} \cite{mohan2019compressed}.

\section{A Brief Overview of Deep Learning}
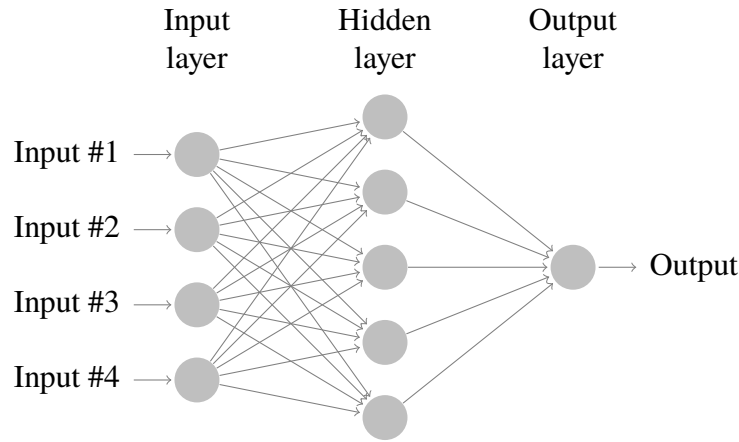
\begin{figure}
\centering

\def\layersep{2.5cm}
\begin{tikzpicture}[shorten >=1pt,->,draw=black!50, node distance=\layersep]
    \tikzstyle{every pin edge}=[<-,shorten <=1pt]
    \tikzstyle{neuron}=[circle,fill=black!25,minimum size=17pt,inner sep=0pt]
    \tikzstyle{input neuron}=[neuron, fill=gray!50];
    \tikzstyle{output neuron}=[neuron, fill=gray!50];
    \tikzstyle{hidden neuron}=[neuron, fill=gray!50];
    \tikzstyle{annot} = [text width=4em, text centered]

    \foreach \name / \y in {1,...,4}
        \node[input neuron, pin=left:Input \#\y] (I-\name) at (0,-\y) {};

    \foreach \name / \y in {1,...,5}
        \path[yshift=0.5cm]
            node[hidden neuron] (H-\name) at (\layersep,-\y cm) {};

    \node[output neuron,pin={[pin edge={->}]right:Output}, right of=H-3] (O) {};

    \foreach \source in {1,...,4}
        \foreach \dest in {1,...,5}
            \path (I-\source) edge (H-\dest);

    \foreach \source in {1,...,5}
        \path (H-\source) edge (O);

    \node[annot,above of=H-1, node distance=1cm] (hl) {Hidden layer};
    \node[annot,left of=hl] {Input layer};
    \node[annot,right of=hl] {Output layer};

\end{tikzpicture}
\caption{Simple neural network representation}
\label{fig:simpleANN}
\end{figure}

Machine learning techniques, especially deep learning using multilayer neural networks and convolutional neural networks\cite{lecun1998gradient}, have been highly successful in addressing a wide variety of computational and data science problems. Unlike rule-based systems, machine learning methods employ the use of features which are mapped from input or user-supplied features, which means that the output is learned and estimated from the input features. In deep learning, the user-supplied features are mapped into more abstract features and the output is learned from these abstract features, thus essentially making deep learning a subset of machine learning methods. Over the last few years, deep learning methods have been extensively applied to computer vision problems, speech recognition, and language translation. Neural networks, which govern how the features are mapped, are a set of procedures that model the input features using hidden units, or neurons, for machine learning. These neural networks are non-linear functions which transform the input data to a parameterized output through a connected sequence of hidden units and these hidden units are stacked together to form hidden layers. Figure \ref{fig:simpleANN} shows a simple neural network with four inputs and one output with one hidden layer containing five hidden units. This is a type of a simple fully connected multi-layer perceptron (MLP) consisting of three layers of nodes- an input layer, a hidden layer, and an output layer. With more hidden layers and more number of such hidden units, the network learns to map more abstract features thus signifying the depth of deep learning methods. 

Unlike the fully connected multi-layer perceptron networks, the convolutional neural networks (CNNs) take into account the spatial distribution of input data which makes them suitable for a variety of physics-based applications including fluid dynamics. In CNNs each layer has detection filters for identifying and learning from different abstract features. In addition to feed-forward networks like MLPs and CNNs, there exists another class of neural networks called recurrent neural networks which also have a feed-back connections to the neurons and the long short-term memory (LSTM)  neural networks belong to this class of feed-back neural networks. These LSTM are well suited for predicting the time series of data which consist of units such as a cell, an input gate, an output gate and a forget gate. The cell keeps a record of values over arbitrary time intervals and the three gates help in regulating the flow of information in and out of the memory cell \cite{hochreiter1997long}. The convolutional neural networks with long short-term memory has been proposed by \cite{xingjian2015convolutional} which can take into account the spatial as well as temporal distribution of the data. %
An activation function is a transfer function which  determines the relation between input and output of a node in neural network, and normalizes the output value to a certain range. This relationship could be linear or non-linear, and depending on the application, a proper choice of activation function for each layer has to be made. For example, a linear or identity activation function is defined as $f(x) = x $, whereas a rectified linear unit (ReLU) activation function is defined as, 

\begin{equation}
f(x)=\left\{
                \begin{array}{ll}
                  0 \quad \textrm{for} \quad x\le0 \\
                  x \quad \textrm{for} \quad x>0 
                \end{array}
              \right.  
\end{equation}

\begin{figure}[t!]
    \centering
\includegraphics[width=0.90\linewidth, height=9cm]{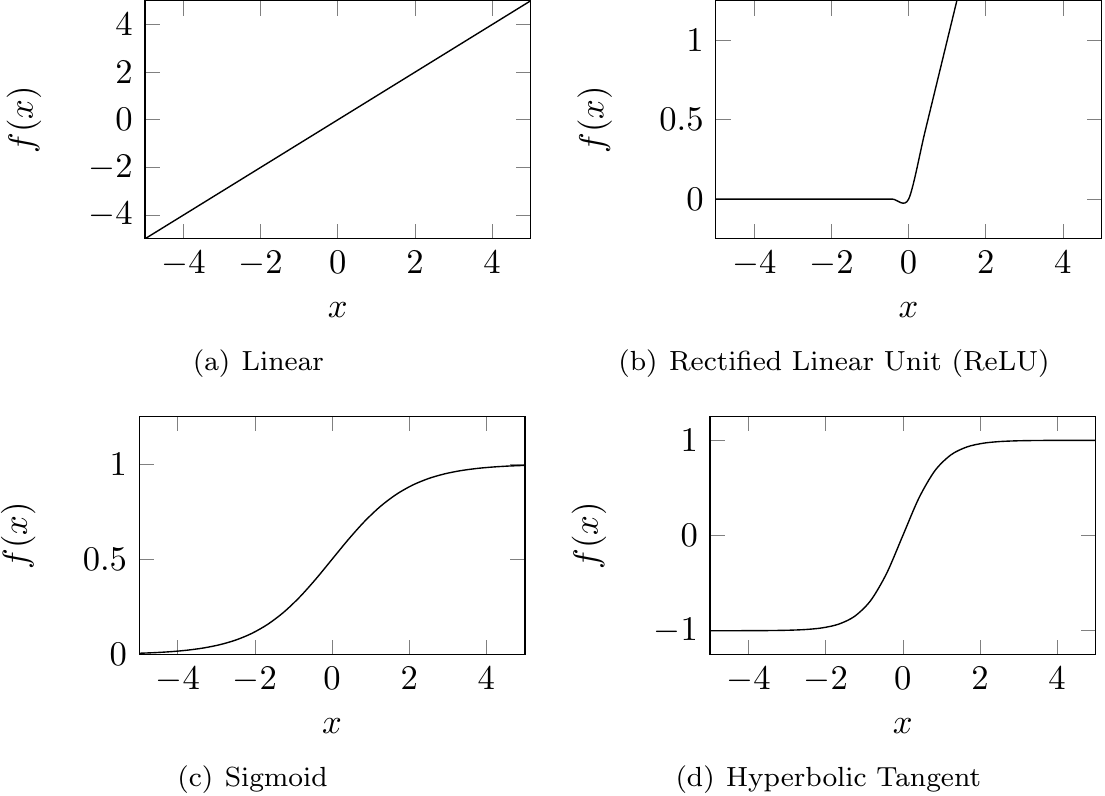}
\caption{Commonly used activation functions.}
\label{fig:activation-functions}
\end{figure}

Some other commonly used activation functions are sigmoid activation function (eq. \ref{eqn:sigmoid}) and hyperbolic tangent activation function (eq. \ref{eqn:tanh}). The input-output relationship between these activation functions is provided in Figure \ref{fig:activation-functions}.
\begin{equation}
f(x) = \sigma(x) = \frac{1}{1 + e^{-x}} 
\label{eqn:sigmoid}
\end{equation}

\begin{equation}
f(x) = tan^{-1}(x) = \frac{e^{x} - e^{-x}}{e^{x} + e^{-x}}
\label{eqn:tanh}
\end{equation}

Depending on dataset as well as the desired application of classification or regression, a suitable activation function has to be chosen. For a detailed overview of deep learning methods and practices, the readers are referred to \cite{goodfellow2016deep} \cite{chollet2018deep}.

\chapter{Generating Inflow Turbulence}

\section{Brief Overview of LES }
\begin{figure}[htbp!]
        \centering
         \includegraphics[width=1.00\linewidth, height=4cm]{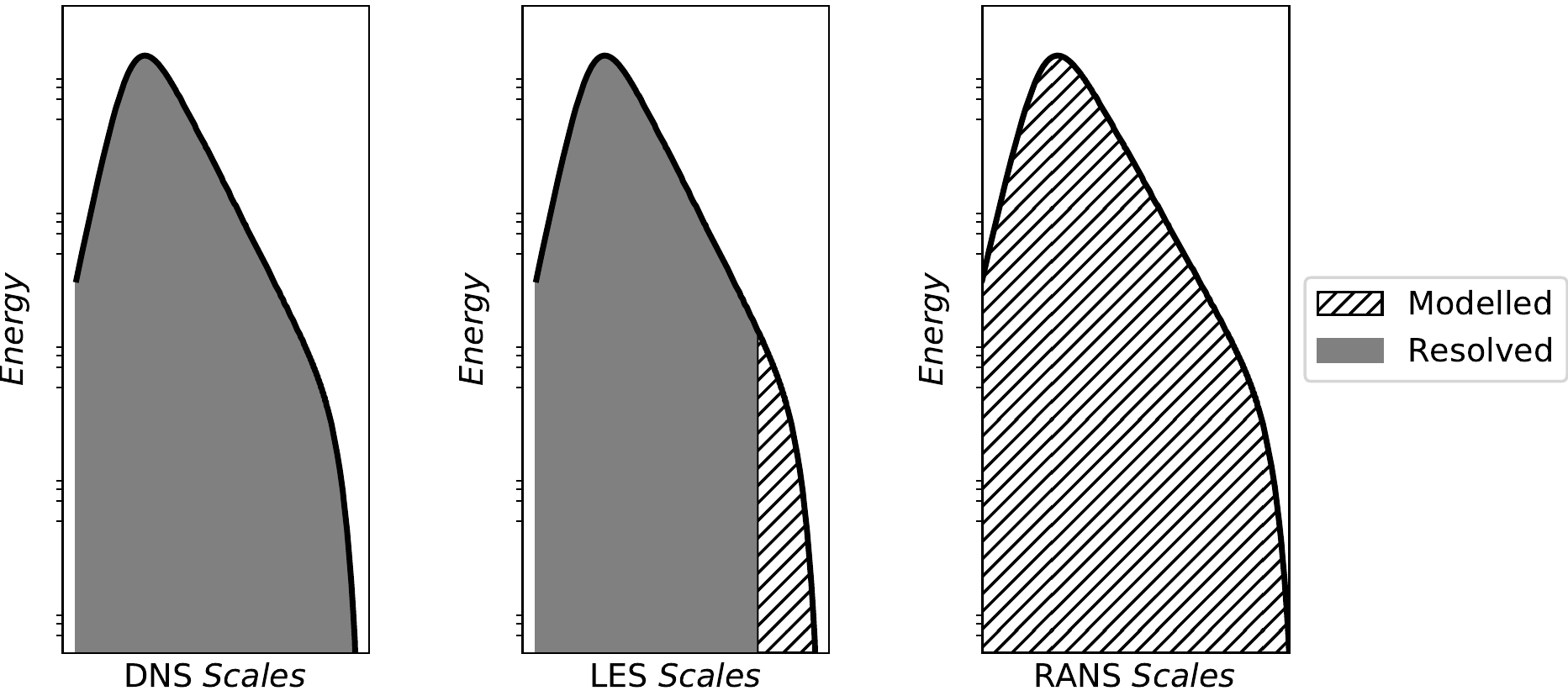} 
         \caption{Representation of resolved and modelled scales of turbulence.}
    \label{fig:dnsLESrans}
    \end{figure}

Based on the accuracy of solving the Navier-Stokes equation, the fluid-simulation methods are broadly grouped into three types - Direct Numerical Simulation (DNS), Large Eddy Simulation (LES), and Reynolds Averaged Navier-Stokes Simulation (RANS). The DNS resolves all the turbulent motions, whereas the LES resolves larger unsteady turbulent motions and models small-scale motions. In terms of accuracy of representing the flow-physics as well as the cost of computation, LES lies between the DNS and RANS. Figure \ref{fig:dnsLESrans} describes the schematics of resolved versus modeled scales on a turbulent kinetic energy spectrum plot. The notion of \textit{scales} can be understood by considering flow-scales as analogous to the size of flow structures, for example, the size of vortices or eddies.   
For LES, the continuity and momentum conservation equations are obtained by filtering the Navier-Stokes equations. Filtering operation involves decomposing the velocity into sum of filtered and residual components which means,
\begin{equation}
u_{i}(x_{i},t) = \underbrace{\widetilde{u_{i}}(x_{i},t)}_{filtered} + \underbrace{u^{\prime}(x_{i},t)}_{residual}
\end{equation}

Note that filtered components are also mentioned as resolved components and the residual components are also mentioned as subgrid-scale(SGS) components. The filtered continuity and momentum conservation equations in conservative form are then written as: 
\begin{equation}
 \frac { \partial \bar{\rho} } { \partial t } + \frac { \partial \left( \bar{\rho} \tilde{u _ { i }} \right) } { \partial x _ { i } } = 0  
\end{equation}
\begin{equation}
 \frac { \partial \left( \bar{\rho} \tilde{u _ { i }} \right) } { \partial t } + \frac { \partial \left( \bar{\rho} \tilde{u _ { i }} \tilde{u _ { j }} \right) } { \partial x _ { j } } = - \frac { \partial \bar{p} } { \partial x _ { i } } + \frac { \partial \tilde{\tau _ { i j }} } { \partial x _ { j } } 
\end{equation}
\begin{equation}
 \bar{\rho} \frac { D \tilde{E} } { D t } = - \frac { \partial \bar{q _ { i }} } { \partial x _ { i } } + \frac { \partial } { \partial x _ { j } } \left( \tilde{\tau _ { i j }} \tilde{u _ { i }} \right) - \frac { \partial } { \partial x _ { i } } \left( \bar{p} \tilde{u _ { i }} \right)  
\end{equation}
\begin{equation}
\tilde{\tau _ { i j }} = -\bar{\rho}(\widetilde{u_{i}u_{j}} - \tilde{u_{i}} . \tilde{u_{j}})
\end{equation}

Here, the difference between filtered product $\widetilde{u_{i}u_{j}}$ and product of filtered velocities $\tilde{u_{i}} . \tilde{u_{j}}$ is defined as the residual stress tensor, which is analogous to Reynolds-stress tensor from the RANS formulation. This residual stress tensor, $\tau_{ij}$, also called as subgrid-scale stress tensor forms the backbone of the LES modeling. 

The Smagorinsky model\cite{smagorinsky1963general}, a type of linear eddy-viscosity model, is the simplest and most widely used LES model in which the anisotropic part of residual stress tensor is written as,
\begin{equation}
\tilde{\tau_{ij}} = -2\nu_{t}\overline{S_{ij}}  
\end{equation}
\begin{equation}
\mbox{  with  }  \nu_{t} = l^{2}_{S_{i j}}\tilde{S} = (C_{S}\Delta)^{2}\tilde{S_{i j}}
\end{equation}
where $\tilde{S_{i j}}$ is the filtered strain-rate, and $l_{s}$ \& $C_{S}$ are Smagorinsky lenghthscale and coefficient respectively. The evaluation of this strain-rate is non-zero at the wall which leads to an overstimation of dissipation near the wall \cite{germano1991dynamic}. Though several approaches have been proposed to handle such a near-wall behaviour, the Wall-Adapting Local Eddy-viscosity (WALE) model \cite{nicoud1999subgrid} makes it convinient to evaluate the strain-rate as well as rotation rates, which both go to zero near the wall.  Another model which also takes care of this near-wall behaviour of strain-rate, in addition to vanishing of $\nu_{t}$ in case of pure shear, is the SIGMA model \cite{nicoud2011sigma}. This model uses $\sigma_{1}$, $\sigma_{2}$, $\sigma_{3}$ as three singular values of the velocity gradient tensor, hence named SIGMA model, where

\begin{equation}
\nu_{t} = (C_{\sigma}\Delta)^{2} D_{\sigma}  
\end{equation}
\begin{equation}
D_{\sigma} = \frac{\sigma_{3}(\sigma_{1}-\sigma_{2}) (\sigma_{2}-\sigma_{3})}{\sigma_{1}^{2}}
\end{equation}

with $C_{\sigma}$=1.5 being the model constant, and $\Delta$ is the characteristic length of the filter and equivalent to the cube root of the node volume.  

The reader is referred to \cite{pope2001turbulent} and \cite{sagaut2006large} for a detailed overview of the LES approaches.

\section{Wall-resolved Large Eddy Simulation }
In LES, there are two different ways to treat the walls, either fully-resolve the near-wall region or perform near-wall modeling with appropriate models and or wall-functions. Wall-resolved simulations are such simulations in which the zone close to the wall, called viscous layer, is fully resolved by the mesh to capture all the small scales with the wall having no-slip conditions. Though this approach is computationally costly, it helps in making an exact computation of the physics near the walls. Also, such a well-resolved simulation would eliminate the need for complex models, thus avoiding additional errors. This viscous layer region is very important in channel flow physics as the production, dissipation, and turbulent kinetic energy all achieve their peak values within 1 $< y^{+} <$ 20  \cite{moser1999direct}.  In the present work, the training data was generated using the wall-resolved large eddy simulation (WRLES) method and the SIGMA model\cite{nicoud2011sigma} was used. 

\section{Grid and Numerics }
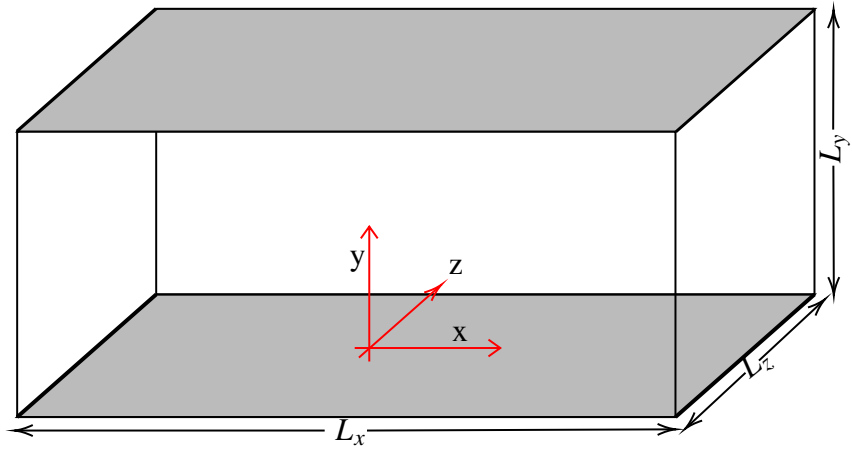
\begin{figure}
\centering
\tikzset{every picture/.style={line width=0.75pt}} 

\begin{tikzpicture}[x=0.75pt,y=0.75pt,yscale=-0.85,xscale=0.85]

\draw  [fill={rgb, 255:red, 187; green, 187; blue, 187 }  ,fill opacity=1 ] (163.1,190.55) -- (553.48,190.55) -- (470.78,263.37) -- (80.4,263.37) -- cycle ;
\draw   (162.89,21) -- (553.48,21) -- (553.48,190.55) -- (162.89,190.55) -- cycle ;
\draw  [line width=0.75]  (80.4,93.82) -- (471,93.82) -- (471,263.37) -- (80.4,263.37) -- cycle ;
\draw [line width=1.5]    (553.48,21) -- (471,93.82) ;

\draw [line width=1.5]    (553.48,190.55) -- (471,263.37) ;

\draw [line width=1.5]    (162.89,21) -- (80.4,93.82) ;

\draw [line width=1.5]    (162.89,190.55) -- (80.4,263.37) ;

\draw [color={rgb, 255:red, 254; green, 0; blue, 0 }  ,draw opacity=1 ] (280.68,222.37) -- (367.05,222.37)(289.32,150.27) -- (289.32,230.38) (360.05,217.37) -- (367.05,222.37) -- (360.05,227.37) (284.32,157.27) -- (289.32,150.27) -- (294.32,157.27)  ;
\draw [color={rgb, 255:red, 255; green, 0; blue, 0 }  ,draw opacity=1 ]   (283.19,228.05) -- (330.8,185.93) ;
\draw [shift={(332.29,184.61)}, rotate = 498.5] [color={rgb, 255:red, 255; green, 0; blue, 0 }  ,draw opacity=1 ][line width=0.75]    (10.93,-3.29) .. controls (6.95,-1.4) and (3.31,-0.3) .. (0,0) .. controls (3.31,0.3) and (6.95,1.4) .. (10.93,3.29)   ;

\draw  [fill={rgb, 255:red, 187; green, 187; blue, 187 }  ,fill opacity=1 ] (163.32,21) -- (553.7,21) -- (471,93.82) -- (80.62,93.82) -- cycle ;
\draw    (287.97,271.38) -- (469.33,270.67) ;
\draw [shift={(471.33,270.67)}, rotate = 539.78] [color={rgb, 255:red, 0; green, 0; blue, 0 }  ][line width=0.75]    (10.93,-3.29) .. controls (6.95,-1.4) and (3.31,-0.3) .. (0,0) .. controls (3.31,0.3) and (6.95,1.4) .. (10.93,3.29)   ;

\draw    (262.86,271.38) -- (80.73,271.38) ;
\draw [shift={(78.73,271.38)}, rotate = 360] [color={rgb, 255:red, 0; green, 0; blue, 0 }  ][line width=0.75]    (10.93,-3.29) .. controls (6.95,-1.4) and (3.31,-0.3) .. (0,0) .. controls (3.31,0.3) and (6.95,1.4) .. (10.93,3.29)   ;

\draw    (564.57,92.68) -- (564.79,23.46) ;
\draw [shift={(564.8,21.46)}, rotate = 450.19] [color={rgb, 255:red, 0; green, 0; blue, 0 }  ][line width=0.75]    (10.93,-3.29) .. controls (6.95,-1.4) and (3.31,-0.3) .. (0,0) .. controls (3.31,0.3) and (6.95,1.4) .. (10.93,3.29)   ;

\draw    (564.69,114.12) -- (564.95,189.23) ;
\draw [shift={(564.96,191.23)}, rotate = 269.8] [color={rgb, 255:red, 0; green, 0; blue, 0 }  ][line width=0.75]    (10.93,-3.29) .. controls (6.95,-1.4) and (3.31,-0.3) .. (0,0) .. controls (3.31,0.3) and (6.95,1.4) .. (10.93,3.29)   ;

\draw    (520.54,229.83) -- (558.41,194.6) ;
\draw [shift={(559.87,193.24)}, rotate = 497.07] [color={rgb, 255:red, 0; green, 0; blue, 0 }  ][line width=0.75]    (10.93,-3.29) .. controls (6.95,-1.4) and (3.31,-0.3) .. (0,0) .. controls (3.31,0.3) and (6.95,1.4) .. (10.93,3.29)   ;

\draw    (512.55,237.08) -- (478.03,269.16) ;
\draw [shift={(476.56,270.52)}, rotate = 317.1] [color={rgb, 255:red, 0; green, 0; blue, 0 }  ][line width=0.75]    (10.93,-3.29) .. controls (6.95,-1.4) and (3.31,-0.3) .. (0,0) .. controls (3.31,0.3) and (6.95,1.4) .. (10.93,3.29)   ;

\draw (343.22,213.76) node  [align=left] {x};
\draw (282.57,170.92) node  [align=left] {y};
\draw (340.72,174.24) node  [align=left] {z};
\draw (278.34,272.23) node   {$L_{x}$};
\draw (565.55,103.82) node [rotate=-270.09]  {$L_{y}$};
\draw (518.85,232.16) node [rotate=-320.73]  {$L_{z}$};

\end{tikzpicture}
\caption{Figure of computational domain and coordinate system in the channel}
\end{figure}

A stretched grid was generated with the HIP package \cite{muller1999coarsening} which is the main tool at CERFACS to manipulate meshes for using with AVBP. The AVBP solver \cite{sch1999steady}, used both for basic and applied research, is an advanced CFD tool for the numerical simulation of unsteady flows. It is an unstructured, explicit, compressible LES solver developed at CERFACS. It relies on the cell-vertex finite-volume method and time marching is done with the Runge-Kutta scheme. A more detailed description of AVBP can be found in the CERFACS internal AVBP manual and in the works of \cite{rudgyard1994modular} \cite{rudgyard1996cpulib}.

In order to obtain a continuous periodic channel flow, a forcing term is introduced in the momentum equation that is applied in the whole computational domain at each time step as shown in red in equation \ref{eqNS_momentum}. %

\begin{equation}
 \frac { \partial \left( \bar{\rho} \tilde{u _ { i }} \right) } { \partial t } + \frac { \partial \left( \bar{\rho} \tilde{u _ { i }} \tilde{u _ { j }} \right) } { \partial x _ { j } } = - \frac { \partial \bar{p} } { \partial x _ { i } } + \frac { \partial \tilde{\tau _ { i j }} } { \partial x _ { j } } \textcolor{red}{ + \mathcal { S } _ { \mathrm { QDM }  } }
\label{eqNS_momentum}
\end{equation}

For the current wall-resolved LES of a turbulent channel flow,  $2^{nd}$ order Lax-Wendroff convection scheme was used with FE 2$\Delta$ discretization for diffusion. CFL number was set to 0.7. The channel's grid is homogeneous and periodic in the streamwise and spanwise directions, and it is stretched with a $tanh$ profile in the wall normal direction. It consists of approximately 52 million grid points with 400$\times$327$\times$400 along $x$, $y$ and $z$ respectively. The grid spacing in wall units is $\Delta$$x^+$= 9.53 and $\Delta$$z^+$ = 4.76, whereas for $y$ direction it is  $\Delta$$y_{wall}^+$ = 1.00 and $\Delta$$y_{center}^+$ = 3.56. Fully developed turbulent statistics were captured for the flow at $Re_{\tau}$ = 950, where  $Re_{\tau}$ is computed as 
\begin{equation}
Re_{\tau} =  \frac{u_{\tau}h}{\nu}
\end{equation}
h is the channel half-height($L_y$/2) and the all quantities are made dimensionless using the friction velocity $u_{\tau}$ as, 
\begin{equation}
y^+ = \frac{y u_{\tau}}{\nu}
\end{equation}
\begin{equation}
U^+ = \frac{U}{u_{\tau}}
\end{equation}
To create a large dataset for deep learning, the present WRLES was run for 15$\tau_{diff}$ after the convergence of statistics. Around 800,000 compute-hours on 65,536 cores of CNRS Turing (IBM Blue Gene/Q) computer were used for this simulation. 

\section{Results}
\begin{figure}[htbp!]
        \centering
         \includegraphics[width=0.85\linewidth, height=7cm]{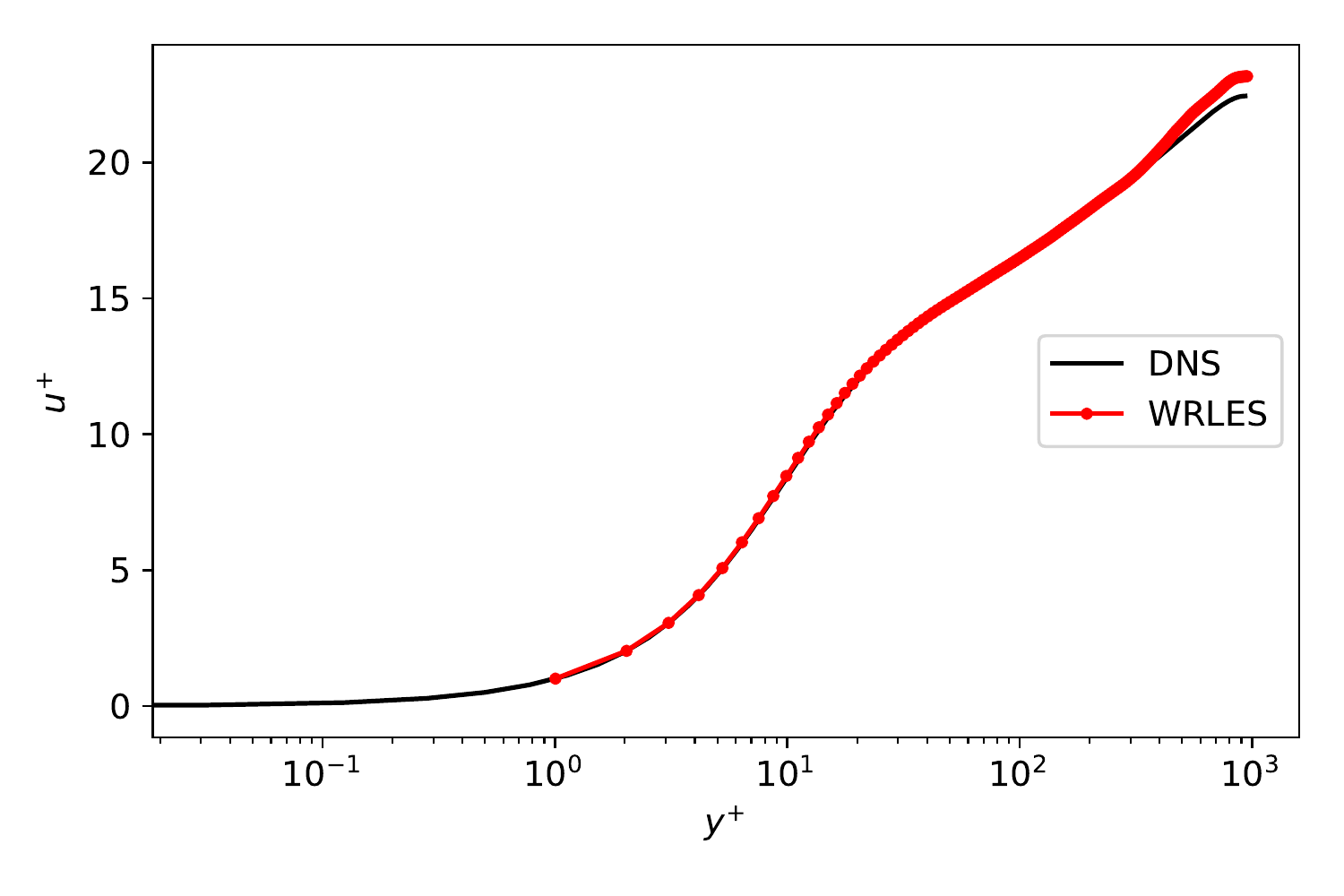} 
         \caption{Normalized mean streamwise velocity profile for current WRLES and DNS}
    \label{fig:wrles_U_profile}
    \end{figure}

Three-dimensional turbulent channel flow is computed by WRLES using the information provided in the previous section. Mean velocity profile for the present result is compared with the DNS\cite{del2003spectra} results in Figure \ref{fig:wrles_U_profile}. The mean velocity profile appears in a good agreement with the DNS data, especially close to the wall.

\begin{figure}[htbp!]
        \centering
         \includegraphics[width=0.85\linewidth, height=7cm]{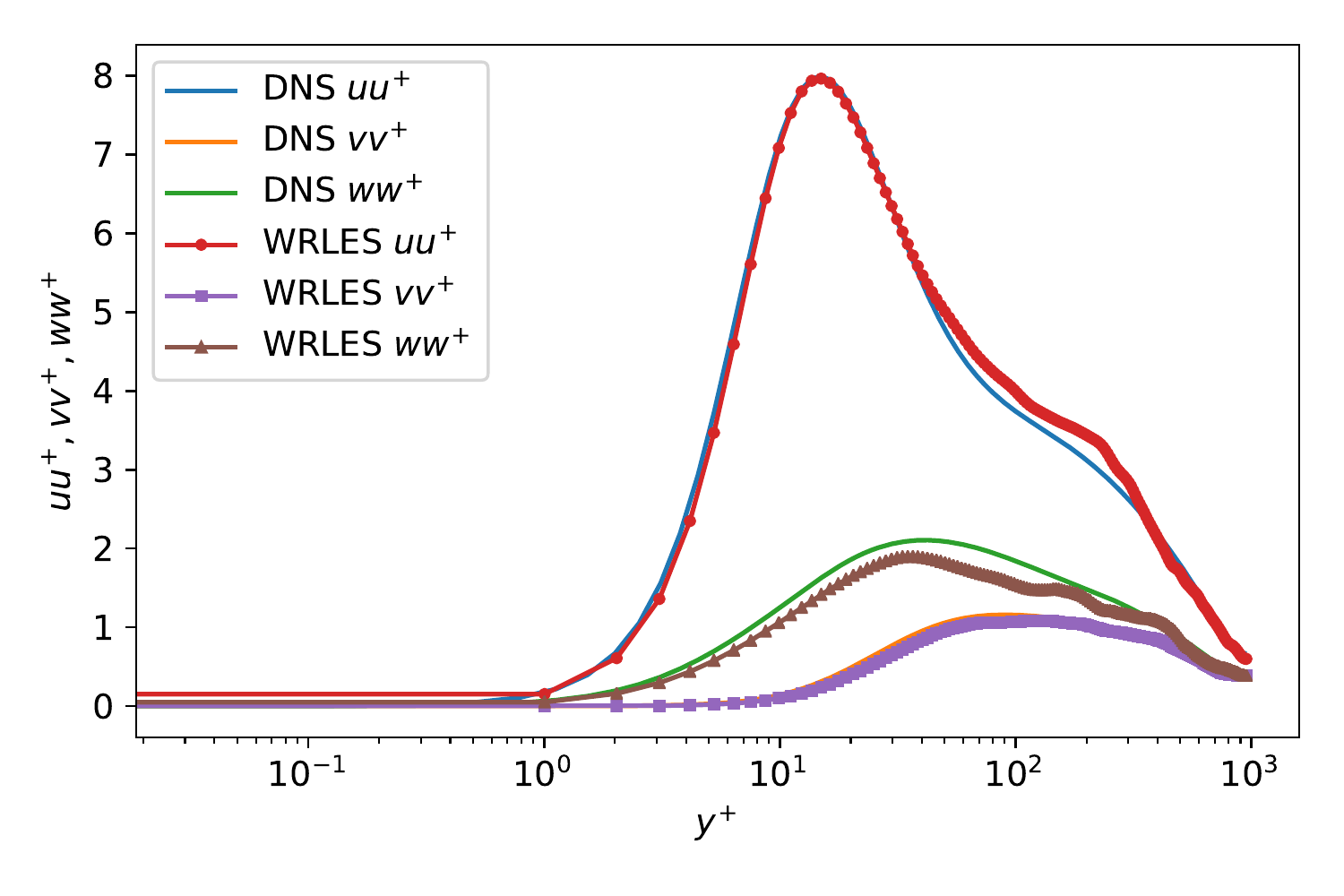} 
         \caption{Normalized Reynolds stresses for current WRLES and DNS}
    \label{fig:wrles_rstrss_profile}
    \end{figure}

Similarly, Reynolds stresses are non-dimensionalised as:
\begin{equation}
u_{i}u_{i}^+ = \frac{u^{\prime}_{i}u^{\prime}_{i}{}^2}{u_{\tau}^2}
\end{equation}
Reynolds stresses from the three velocity components are compared as functions of the wall-normal coordinate in wall units, from top to bottom: streamwise($uu^+$), spanwise($ww^+$), and wall-normal component($vv^+$). Solid lines show DNS at $Re_{\tau}$=950 \cite{del2003spectra}, dotted lines show the present WRLES results. These results are in good agreement with the DNS results as shown in Figure \ref{fig:wrles_rstrss_profile}. 

Additonal paramters of the WRLES are provided in the Table \ref{tableWRLESParams}.
\begin{table}[htpb!]
\begin{center}
\begin{tabular}{|l|l|}
\hline
\textit{Parameter} & \textit{Value} \\ \hline
Bulk Reynolds number  &  77240 \\
Friction Reynolds number  &  952.449\\
Mach &  0.09617\\
Pressure &  5.00E+05 $Pa$\\ 
Bulk velocity &  34.530 $m/s$\\
Friction Velocity $u_{\tau}$ &  1.699\\
Channel half-height $h$  &  0.0020 $m$ \\
Domain size (in wall units)  &  $Lx^+$ = 3801, \\ 
 & $Ly^+$ = 1901, \\ 
 & $Lz^+$ = 1901 \\
Computational time &  15$\tau_{diff}$ \\ 
Diffussion time $\tau_{diff}$ & 0.00117693 $s$ \\
Convective time  &  2.316801E-04 $s$ \\
Computational time-step $\Delta t$  &  1.545E-08 $s$\\
\hline
\end{tabular}
  \caption{{Details of the WRLES Simulation}}
  \label{tableWRLESParams}
\end{center}
\end{table}

\begin{figure}[htbp!]
        \centering
         \includegraphics[width=0.95\linewidth, height=5cm]{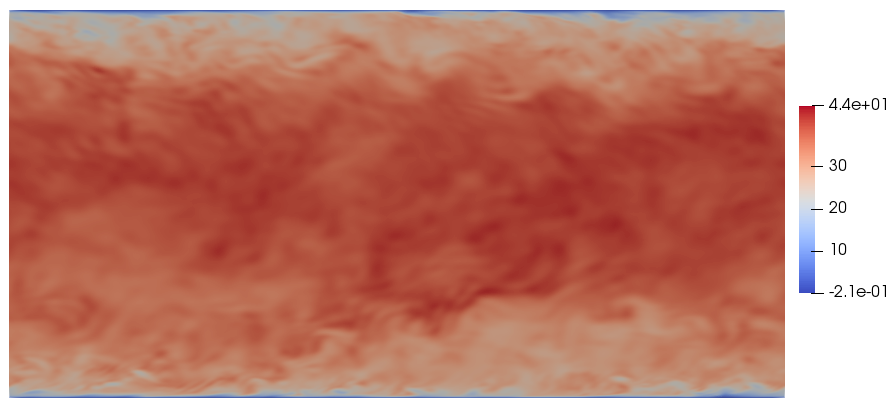} 
         \caption{$u$ velocity snapshot along plane $x-y$ for current WRLES }
    \label{fig:wrles_snapshot_stream}
    \end{figure}

\begin{figure}[htbp!]
        \centering
         \includegraphics[width=0.45\linewidth, height=5cm]{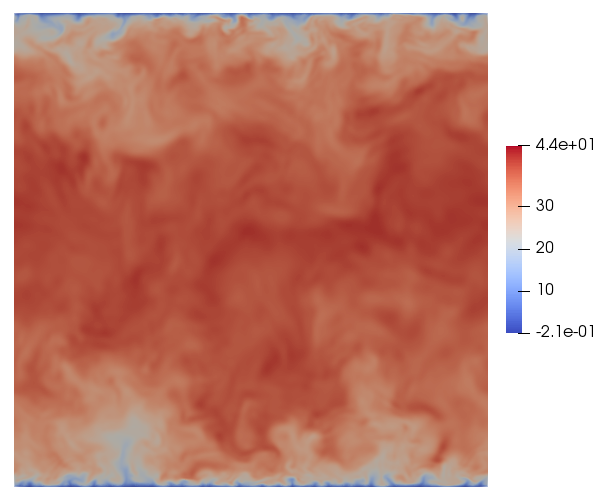} 
         \caption{$u$ velocity snapshot along plane $y-z$ for current WRLES }
    \label{fig:wrles_snapshot_span}
    \end{figure}

Figure \ref{fig:wrles_snapshot_stream} shows the visualization of streamwise velocity along the $x-y$ plane, whereas figure (\ref{fig:wrles_snapshot_span}) shows the visualization of a snapshot of spanwise velocity along the $y-z$ plane.

\section{Data Collection and Preparation for Learning}
\begin{figure}
\centering

\tikzset{every picture/.style={line width=0.75pt}} 

\begin{tikzpicture}[x=0.75pt,y=0.75pt,yscale=-0.65,xscale=0.65]

\draw   (140.55,8) -- (607.24,8) -- (607.24,205.74) -- (140.55,205.74) -- cycle ;
\draw  [fill={rgb, 255:red, 187; green, 187; blue, 187 }  ,fill opacity=1 ] (140.81,205.74) -- (607.24,205.74) -- (508.43,290.67) -- (42,290.67) -- cycle ;
\draw  [fill={rgb, 255:red, 240; green, 111; blue, 111 }  ,fill opacity=0.94 ] (368.81,7.33) -- (368.55,205.08) -- (270,290) -- (270.5,170.67) -- (270.26,92.26) -- cycle ;
\draw  [line width=0.75]  (42,92.92) -- (508.69,92.92) -- (508.69,290.67) -- (42,290.67) -- cycle ;
\draw [line width=1.5]    (607.24,8) -- (508.69,92.92) ;

\draw [line width=1.5]    (607.24,205.74) -- (508.69,290.67) ;

\draw [line width=1.5]    (140.55,8) -- (42,92.92) ;

\draw [line width=1.5]    (140.55,205.74) -- (42,290.67) ;

\draw [line width=1.5]    (368.81,7.33) -- (368.55,205.08) ;

\draw  [fill={rgb, 255:red, 187; green, 187; blue, 187 }  ,fill opacity=0.72 ] (141.07,8) -- (607.5,8) -- (508.69,92.92) -- (42.26,92.92) -- cycle ;
\draw [line width=1.5]    (270.26,92.26) -- (270,290) ;

\draw [line width=1.5]    (368.55,205.08) -- (270,290) ;

\draw [line width=1.5]    (368.81,7.33) -- (270.26,92.26) ;

\end{tikzpicture}
\caption{Plane at which temporal data was collected in the channel}
\label{fig:channelCutx}
\end{figure}
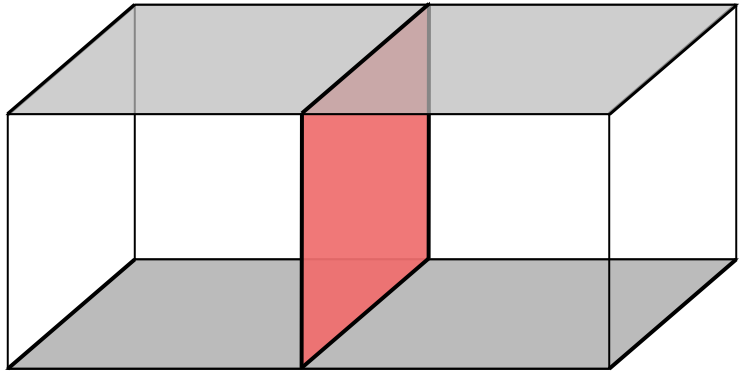

Full fields after convergence were collected for computing mean statistics of the channel flow. Primarily, temporal snapshots were collected at a streamwise plane $x= Lx / 2$ , thus in the 2D plane $(y-z)$. Figure \ref{fig:channelCutx} shows the plane at which this temporal data was collected where the fields $u$, $v$, $w$, and $P$ were stored. The results produced by AVBP were on an unstructured grid, hence the data of these snapshots was converted from unstructured HDF5 files to structured data in $.npy$ which is a standard binary file format in NumPy. Though this involved significant data-processing efforts, it proved beneficial later in the deep learning training procedure. Data of these snapshots were stored after every 10 iterations, the choice of which was based on the available memory and the memory requirement of each snapshot. These 10 iterations correspond to a minor visual change in the distribution of field parameters, thus it represents a $movie$ of temporal evaluation of field parameters such as velocity components and pressure. In terms of physical time, 10 iterations correspond to nearly 200 nanoseconds, whereas in terms of diffusion times, it corresponds to nearly 1/5000 $^{th}$ of one diffusion time. %

\chapter{Learning Turbulent Fields }

\section{Distribution of Data}
\begin{figure}[htbp!]
        \centering
         \includegraphics[width=0.8\linewidth, height=7cm]{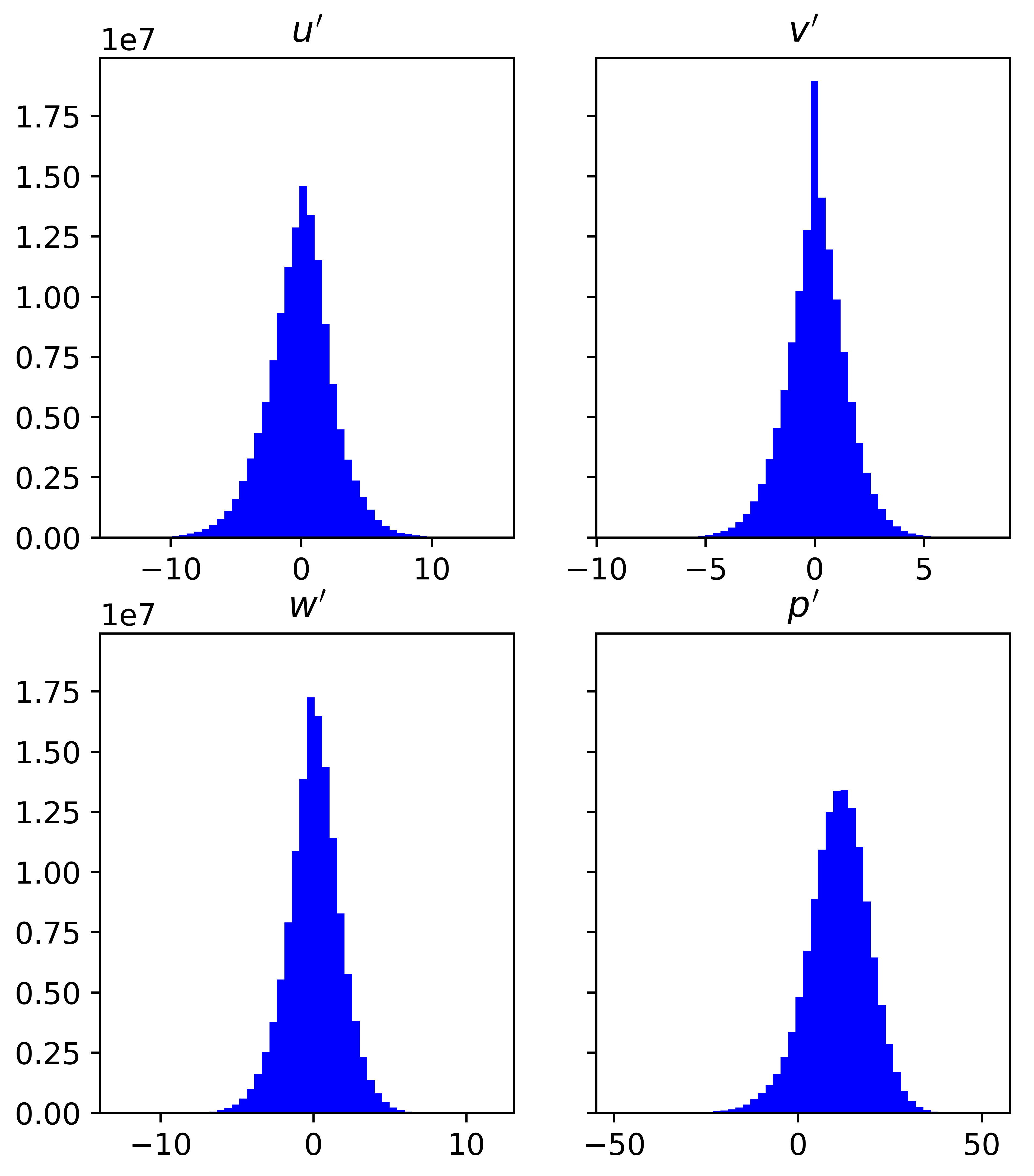} 
         \caption{Distribution of fluctuations for current WRLES }
    \label{plot_dataHistogram}
    \end{figure}

Since the objective was to replicate a $movie$ of the evolution of turbulent flow-fields, fluctuations were computed from the snapshot data of the velocity and pressure fields as:
\begin{equation}
u_i^{\prime} = u_i - \overline{u_{i,t}} 
\end{equation}
\begin{equation}
p^{\prime} = p - \overline{p_{t}} 
\end{equation}
This was achieved by computing a mean snapshot in time and subtracting this mean snapshot from every other snapshot in time. Figure \ref{plot_dataHistogram} shows the total distribution of samples of these fluctuations. The distribution is spread in different ranges for each flow-field. It was observed that the distribution of the streamwise velocity component($u^{\prime}$) for training dataset is skewed due to the mean flow in the streamwise direction. It has been shown by \cite{shanker1996effect} that such an uneven distribution of input data to neural networks adversely affects the training and thus prediction. Also, if the mean values of inputs deviate from zero, the weight update step in the deep learning training could get affected, making the task of learning slower\cite{cun1998efficient}. To avoid these complications, the fluctuations were standardized such that their mean value is zero and the standard deviation is one. While making predictions, the output data was again de-standardized so that actual input fluctuations and the predicted output fluctuations could be compared.   

\section{Training and Evaluation Method}
As it is customary in machine learning, the snapshot dataset was split into a training (75\%) and a testing (25\%) dataset. The weights of the neural network are tuned in a supervised manner to fit the training dataset. In order to monitor the overfitting of this data, accuracy is periodically evaluated on the test set. These datasets consist of pairs of input vectors and the corresponding output vectors. 

In the current work the input vectors $q^{t}$ consist of the standardized fluctuations $u_{i}^{\prime}$ and $p^{\prime}$ at time $t$ , and the output vectors $q^{t+1}$ consist of same fields but at time $t+1$. Note that the difference between $t$ and $t+1$ correspond to $10$ time-steps in the wall-resolved LES as described in the previous chapter.  Thus, the deep learning model that we are trying to learn can be represented with $\boldsymbol{q} = [u^{\prime} v^{\prime} w^{\prime} p^{\prime} ]$ as:
\begin{equation}
\boldsymbol {q}^{t+1}_{\mathrm Learned}
={\mathcal {F}}({\boldsymbol {q}^{t}_{\mathrm LES}}; {\boldsymbol {W}})\approx \boldsymbol {q}^{t+1}_{\mathrm LES}
\end{equation}
Here the training procedure would optimize the weights $\boldsymbol {W}$ and a non-linear mapping function $\mathcal {F}()$  would be learned.
This learning is achieved by minimizing the loss function which, in the current study, was the mean-squared error between ${q}^{t+1}_{\mathrm Learned}$ and ${q}^{t+1}_{\mathrm LES}$.

After the deep learning model is trained, the learned model was used to perform an \textit{a priori} simulation of turbulence generation. This was achieved by first using the trained model to make a prediction from an input WRLES snapshot and then recycling this predicted output to the input of the same model, and continuing this recursively. 
\newpage 
This \textit{a priori} simulation can be represented as:
\begin{gather*} 
\boldsymbol {q}^{t=1}_{\mathrm Learned}
={\mathcal{F}}({\boldsymbol {q}^{t=0}_{\mathrm LES}}; {\boldsymbol {W_{Learned}}}) \\
\boldsymbol {q}^{t=2}_{\mathrm Learned}
={\mathcal{F}}({\boldsymbol {q}^{t=1}_{\mathrm Learned}}; {\boldsymbol {W_{Learned}}})\\
\boldsymbol {q}^{t=3}_{\mathrm Learned}
={\mathcal{F}}({\boldsymbol {q}^{t=2}_{\mathrm Learned}}; {\boldsymbol {W_{Learned}}})\\
\boldsymbol {q}^{t=N_{iter}}_{\mathrm Learned}
={\mathcal{F}}({\boldsymbol {q}^{t=N_{iter}-1}_{\mathrm Learned}}; {\boldsymbol {W_{Learned}}})
\end{gather*}
Here, the $N_{iter}$ should ideally be large enough to compute turbulence statistics, for example of the order of 12 to 15 times $\tau_{diffussion}$ which roughly corresponds to about 8000 iterations based on the input data. But in the present study, several issues were observed which lead to divergence and eventual crash of the \textit{a priori} simulation. It was observed that the values start diverging near the wall. 
Several cutoff filters on mean-statistics were tested to avoid such divergence, but the results presented in the following sections are without any filtering. Thus, the \textit{a priori} statistics for each of the cases described in the next sections have been collected until different $N_{iter}$. It has to be noted that when the simulation diverges, it could be restarted several times to collect the mean statistics of these simulations until divergence is observed and this was done for Case-2 in the present work. Table \ref{tableCases} shows the cases with their $N_{iter}$ and the type of deep-learning architecture used.

\section{Deep Learning Model Architecture}
\begin{figure}[!htbp]
\tikzset{every picture/.style={line width=0.75pt}} 

\begin{tikzpicture}[x=0.75pt,y=0.75pt,yscale=-0.75,xscale=0.85]

\draw   (238.01,160.72) -- (101.5,194) -- (102.82,13.71) -- (238.84,48.04) -- cycle ;
\draw   (421.92,43.62) -- (561.5,10.35) -- (560.18,190.64) -- (421.09,156.3) -- cycle ;
\draw   (280.09,156.07) -- (280.85,53.05) -- (379.85,53.43) -- (379.09,156.45) -- cycle ;
\draw   (59.5,93.75) -- (84.7,93.75) -- (84.7,87) -- (101.5,100.5) -- (84.7,114) -- (84.7,107.25) -- (59.5,107.25) -- cycle ;
\draw   (238.5,93.75) -- (263.7,93.75) -- (263.7,87) -- (280.5,100.5) -- (263.7,114) -- (263.7,107.25) -- (238.5,107.25) -- cycle ;
\draw   (379.5,93.75) -- (404.7,93.75) -- (404.7,87) -- (421.5,100.5) -- (404.7,114) -- (404.7,107.25) -- (379.5,107.25) -- cycle ;
\draw   (560.5,93.75) -- (585.7,93.75) -- (585.7,87) -- (602.5,100.5) -- (585.7,114) -- (585.7,107.25) -- (560.5,107.25) -- cycle ;

\draw (172.73,98.12) node   {$\mathcal{F}_{encoder}$};
\draw (495.27,99.23) node   {$\mathcal{F}_{decoder}$};
\draw (335.48,100.75) node   {$\mathcal{F}_{latent-space}$};
\draw (38,100) node   {$q^{t}$};
\draw (627,100) node   {$q^{t+1}$};
\draw (320,218) node   {$q^{t+1} =\mathcal{F}_{decoder} \ \left(\mathcal{F}_{latent-space}\left(\mathcal{F}_{encoder}\left( q^{t}\right)\right)\right)$};

\end{tikzpicture}
\caption{Representation of operations in the autoencoder}
\label{figAutoencoder}
\end{figure}
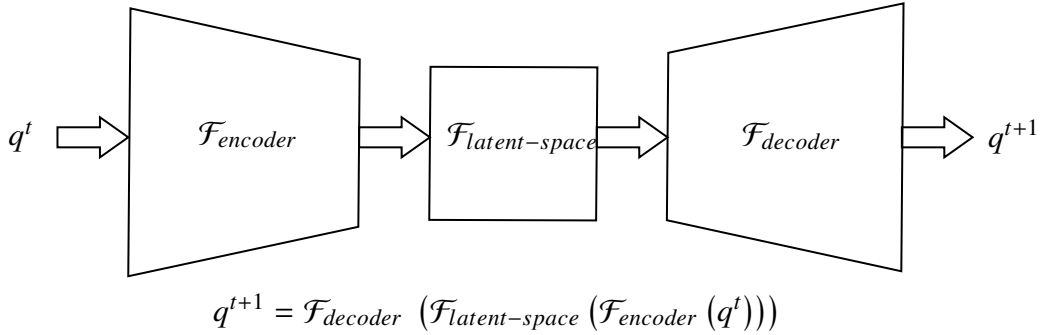

The neural network architectures chosen here are convolutional neural networks (CNN) i.e. stacks if convolutional layers, organized as an \textit{autoencoder}
\cite{hinton2006reducing}. Autoencoders have two parts - a converging part that decreases the spatial dimension of the input (the encoder), and a diverging part that rebuilds an output of the same size as input (the decoder). These two elements are connected by a \textit{latent vector}, which holds the compressed view of the input. In the present work, two different types of methods were used to learn and modify the latent space: multi-layer perceptron (MLP) and long short-term memory (LSTM). The encoder and decoder handle %
the spatial-dimensionality reduction by compressing the high-dimensional spatial data to low-dimensional latent space data, whereas MLPs and LSTMs handle the temporal history preservation. 
Figure \ref{figAutoencoder} shows a representation of operations involved in an autoencoder. The use of such deep learning models for studies around turbulence generation has been recently demonstrated by \cite{fukami2019synthetic} and \cite{mohan2019compressed}.   

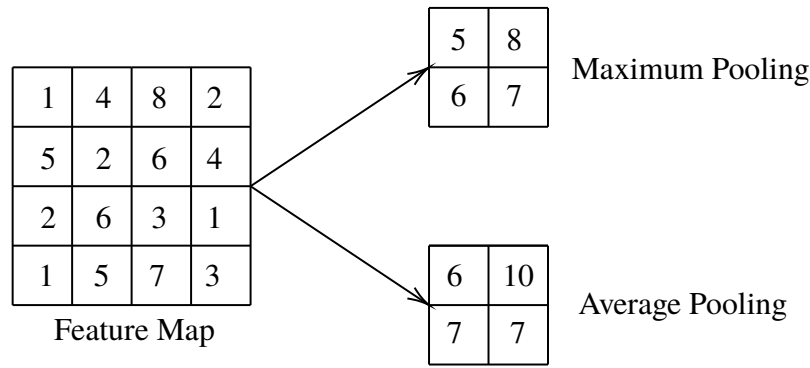
\begin{figure}
\centering
\tikzset{every picture/.style={line width=0.75pt}} 

\begin{tikzpicture}[x=0.75pt,y=0.75pt,yscale=-1,xscale=1]

\draw  [draw opacity=0][line width=0.75]  (153,39) -- (213,39) -- (213,99) -- (153,99) -- cycle ; \draw  [line width=0.75]  (153,39) -- (153,99)(183,39) -- (183,99) ; \draw  [line width=0.75]  (153,39) -- (213,39)(153,69) -- (213,69) ; \draw  [line width=0.75]   ;
\draw  [draw opacity=0][line width=0.75]  (153,99) -- (213,99) -- (213,159) -- (153,159) -- cycle ; \draw  [line width=0.75]  (153,99) -- (153,159)(183,99) -- (183,159) ; \draw  [line width=0.75]  (153,99) -- (213,99)(153,129) -- (213,129) ; \draw  [line width=0.75]   ;
\draw  [draw opacity=0][line width=0.75]  (213,39) -- (273,39) -- (273,99) -- (213,99) -- cycle ; \draw  [line width=0.75]  (213,39) -- (213,99)(243,39) -- (243,99) ; \draw  [line width=0.75]  (213,39) -- (273,39)(213,69) -- (273,69) ; \draw  [line width=0.75]   ;
\draw  [draw opacity=0][line width=0.75]  (213,99) -- (273,99) -- (273,159) -- (213,159) -- cycle ; \draw  [line width=0.75]  (213,99) -- (213,159)(243,99) -- (243,159) ; \draw  [line width=0.75]  (213,99) -- (273,99)(213,129) -- (273,129) ; \draw  [line width=0.75]   ;
\draw    (153,159) -- (273,159) ;

\draw    (273,39) -- (273,159) ;

\draw  [draw opacity=0][line width=0.75]  (363,9) -- (423,9) -- (423,69) -- (363,69) -- cycle ; \draw  [line width=0.75]  (363,9) -- (363,69)(393,9) -- (393,69) ; \draw  [line width=0.75]  (363,9) -- (423,9)(363,39) -- (423,39) ; \draw  [line width=0.75]   ;
\draw    (363,69) -- (423,69) ;

\draw    (423,69) -- (423,9) ;

\draw  [draw opacity=0][line width=0.75]  (363,129) -- (423,129) -- (423,189) -- (363,189) -- cycle ; \draw  [line width=0.75]  (363,129) -- (363,189)(393,129) -- (393,189) ; \draw  [line width=0.75]  (363,129) -- (423,129)(363,159) -- (423,159) ; \draw  [line width=0.75]   ;
\draw    (363,189) -- (423,189) ;

\draw    (423,189) -- (423,129) ;

\draw    (363,129) -- (423,129) ;

\draw    (273,99) -- (361.34,40.11) ;
\draw [shift={(363,39)}, rotate = 506.31] [color={rgb, 255:red, 0; green, 0; blue, 0 }  ][line width=0.75]    (10.93,-3.29) .. controls (6.95,-1.4) and (3.31,-0.3) .. (0,0) .. controls (3.31,0.3) and (6.95,1.4) .. (10.93,3.29)   ;

\draw    (273,99) -- (361.34,157.89) ;
\draw [shift={(363,159)}, rotate = 213.69] [color={rgb, 255:red, 0; green, 0; blue, 0 }  ][line width=0.75]    (10.93,-3.29) .. controls (6.95,-1.4) and (3.31,-0.3) .. (0,0) .. controls (3.31,0.3) and (6.95,1.4) .. (10.93,3.29)   ;

\draw (213,54) node  [align=left] {1 \ \ \ \ 4 \ \ \ \ 8 \ \ \ \ 2};
\draw (213,85) node  [align=left] {5 \ \ \ \ 2 \ \ \ \ 6 \ \ \ \ 4};
\draw (213,115) node  [align=left] {2 \ \ \ \ 6 \ \ \ \ 3 \ \ \ \ 1};
\draw (212,144) node  [align=left] {1 \ \ \ \ 5 \ \ \ \ 7 \ \ \ \ 3};
\draw (398,25) node  [align=left] {5 \ \ \ \ 8 \ \ };
\draw (396,53) node  [align=left] {6 \ \ \ \ 7 \ };
\draw (400,144) node  [align=left] {6 \ \ \ \ 10 \ \ };
\draw (396,173) node  [align=left] {7 \ \ \ \ \ 7 \ };
\draw (214.5,173) node  [align=left] {Feature Map};
\draw (491,160) node  [align=left] {Average Pooling};
\draw (495,41) node  [align=left] {Maximum Pooling};

\end{tikzpicture}
\caption{A representation of convolution with max pooling and average pooling operations.}
\label{fig:avgMaxPool}
\end{figure}

The dimension of each input snapshot data in the present work was $328 \times 400 \times 4 $ i.e. the $y-z$ snapshot-plane with dimension $328 \times 400$ and $u_i^{\prime}$, $p^{\prime}$ were 4 $channels$ put together as $q$. This means the total number of inputs at every time instant is  $524800$.  This input data is compressed with successive convolutions into a low-dimensional compressed feature-map which is the latent-space. The method of pooling helps in reducing the dimensionality of a feature map. For example, a $N_{y} \times N_{z}$ feature map can be reduced to $N_y$/2 $\times$ $N_z$/2 using a pooling layer with a size of 2 i.e. reducing by half. An essential aspect of such pooling is that it should be able to preserve the most important feature of the map, hence there exist methods like max-pooling and average pooling. In max-pooling, the largest value from the feature window is used, whereas in average pooling, an average value of the values in the feature window is used. Figure (\ref{fig:avgMaxPool}) shows a representation of average vs max pooling. It has been shown that for classification tasks, max-pooling operation gives better results \cite{boureau2010theoretical}, and for regression tasks, average pooling yields better results. Since the current work is a regression task, average pooling was used. Strided convolution is another method which can help in reducing the spatial dimensionality. Striding is achieved by skipping a $N_s$ number of elements while moving the convolution window. Thus, if $N_s$=2, the spatial convolution reduces the dimensions by half. 

In the latent-space, a mapping is learned between the successive input snapshots to provide the temporal evolution of predictions. 
Two different cases are presented in the current work: Case-1 uses a fully-connected MLP in the latent space; Case-2 uses a CNN-LSTM\cite{xingjian2015convolutional}, as mentioned in Table(\ref{tableCases}).  After the temporal learning, the low-dimensional latent space data is upsampled into the original dimensions. This upsampling can either be performed with a nearest-neighbor interpolation or with an inverse convolution operation. Though upsampling with interpolation is computationally cheap, it doesn't offer any real advantage in terms of learning. Whereas, in the inverse convolutions, the kernel is learned while training the model just like a CNN. In the present work, inverse convolutions were performed to upsample the low-dimensional data to the original field dimensions. 

Since CNNs come from and were mostly used in the image-classification community, there was no need for physically realistic boundary conditions. The CNNs can have boundary conditions in the form of padding. In the classical sense, padding is used to preserve the spatial dimensions of the field being convoluted, but padding with zeros everywhere doesn't represent the correct physical behavior. To preserve the boundary conditions for the dimensions after successive convolutions, padding with zeros everywhere may violate the notion of $walls$ for the channel flow. To address this issue, a boundary condition formulation was implemented for CNNs such that the walls could be padded with zeros if required and the periodic sides could be padded with values from the periodic cells. This modification has considerably improved the outcomes near the wall region.

The convolutional layers used in the present work rely on 3$\times$3 filters. For Case-1 2$\times$2 pooling is used and for Case-2, strided convolutions with $N_s$=2 are used. The activation functions are the rectified linear units ($ReLU$) for all the layers, and a linear activation is used at the end of the last layer. To optimize the weights of the deep learning model during training, the adaptive momentum estimation optimizer, popularly known as Adam \cite{kingma2014adam}, was used. Adam optimizer uses the classical stochastic gradient descent procedure to update network weights iteratively. The Table \ref{tableCases} shows the details of the two cases presented in this study.

\begin{table}[t!]
\begin{center}
\begin{tabular}{ccccc}
Case & Latent-Space & Epochs & MSE & \textit{a Priori} $N_{iter}$ \\ \hline
Case-1 & MLP & 125 & 0.19 & 3000 \\ 
Case-2 & CNN-LSTM & 250 & 0.02 & 500 \\

\hline
\end{tabular}
  \caption{{Details of cases under investigation}}
  \label{tableCases}
\end{center}
\end{table}

The full training dataset is shown repeatedly to the network during the training, and each pass is referred to as an \textit{epoch}. 
In the current work, an early stopping criterion was used along with a reduction of learning rate if learning doesn't improve after every 35 epochs. An important point to be noted here is that the mean squared error (MSE) is not a full measure of the error in the \textit{a priori} simulation, but it is the error during the training. This means that even a very small error in training could lead to divergence while performing \textit{a priori} simulation later. The implementation of these deep learning methods was done in Python 3.6 using the Keras library \cite{chollet2015keras} which runs on top of TensorFlow\cite{abadi2016tensorflow}. Computationally heavy training of deep learning models, as well as the \textit{a priori} simulation, was done on an Nvidia Tesla V100 GPU.

\subsection{Case-1}
\begin{figure}[htbp!]
        \centering
         \includegraphics[width=0.99\linewidth, height=5cm]{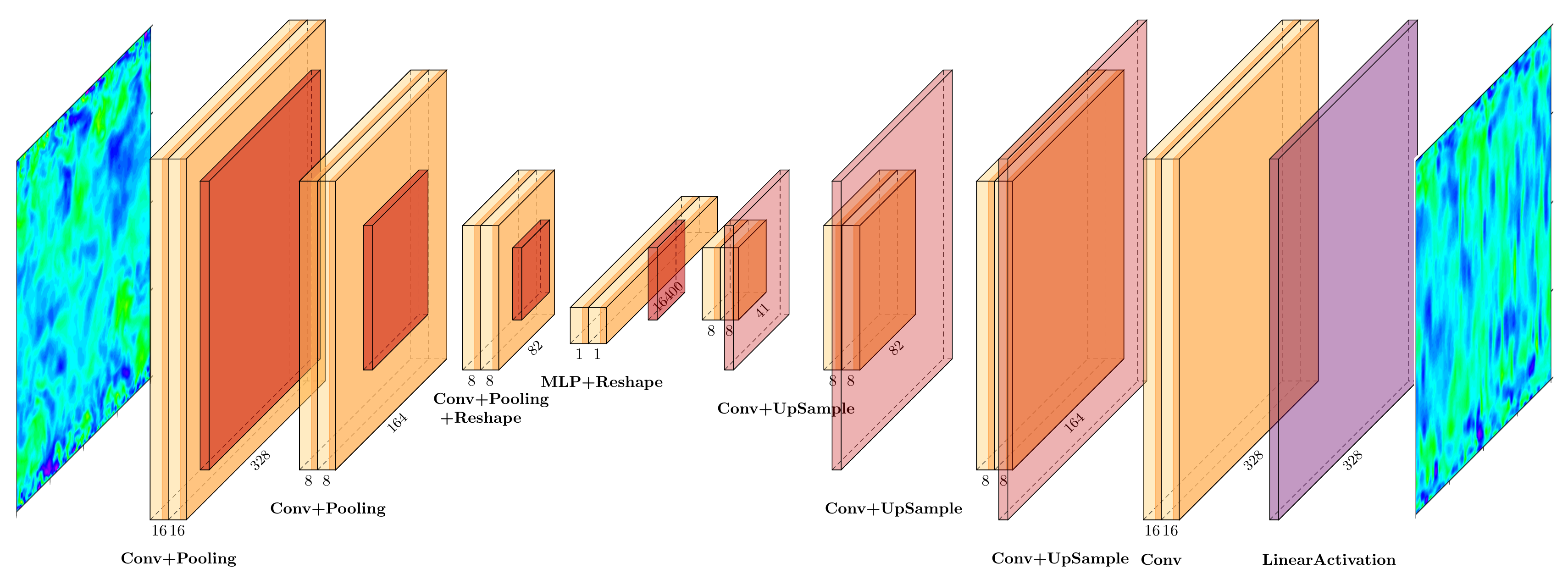} 
         \caption{Deep learning architecture for Case-1 }
    \label{Case1ArchFig}
    \end{figure}

In Case-1, three successive convolutional neural networks along with a pooling layer after every convolution block were used to reduce the spatial dimensions of input data from  $328 \times 400 $ to $41 \times 50 $, and after the latent-space operation, the same compressed spatial data of  $41 \times 50$ was then successively upsampled back to its original dimensions of $328 \times 400$ using a simple resizing with nearest-neighbor interpolation along with successive intermediate convolutional blocks. For temporal learning, a fully connected multi-layer perceptron model was used in the latent-space with the shape of input as well as output as $41 \times 50$. A simplified representation of this model is shown in Figure \ref{Case1ArchFig} and the detailed architecture of this neural network can be found in Table \ref{Case1Arch}. Recalling the introduction of deep learning, it can be noted that the number of $filters$ is an important parameter in describing the $width$ of a deep learning model. The more the number of filters, the more abstract features the neural network could learn. In the current case, the number of filters start with $16$, thus the $4$ input channels are mapped to $16$ channels in the CNN making the total dimension of the first layer as $328 \times 400 \times 16$. After successive convolutional layers, latent space dimension becomes $41 \times 50 \times 8$. This implies that the dimension of the fully connected  MLP is $16400$, and two such successive MLPs are used. Before upsampling, the MLP output is reshaped to $41 \times 50 \times 8$ which was the shape of its input and then the upsampling along with successive convolutions are performed with the same size of filters as that used while dimension reduction. At last, a convolution operation along with a linear activation and $4$ filters produces the output with the same dimensions as that of input. 

\subsection{Case 2}
\begin{figure}[htbp!]
        \centering
         \includegraphics[width=0.99\linewidth, height=5cm]{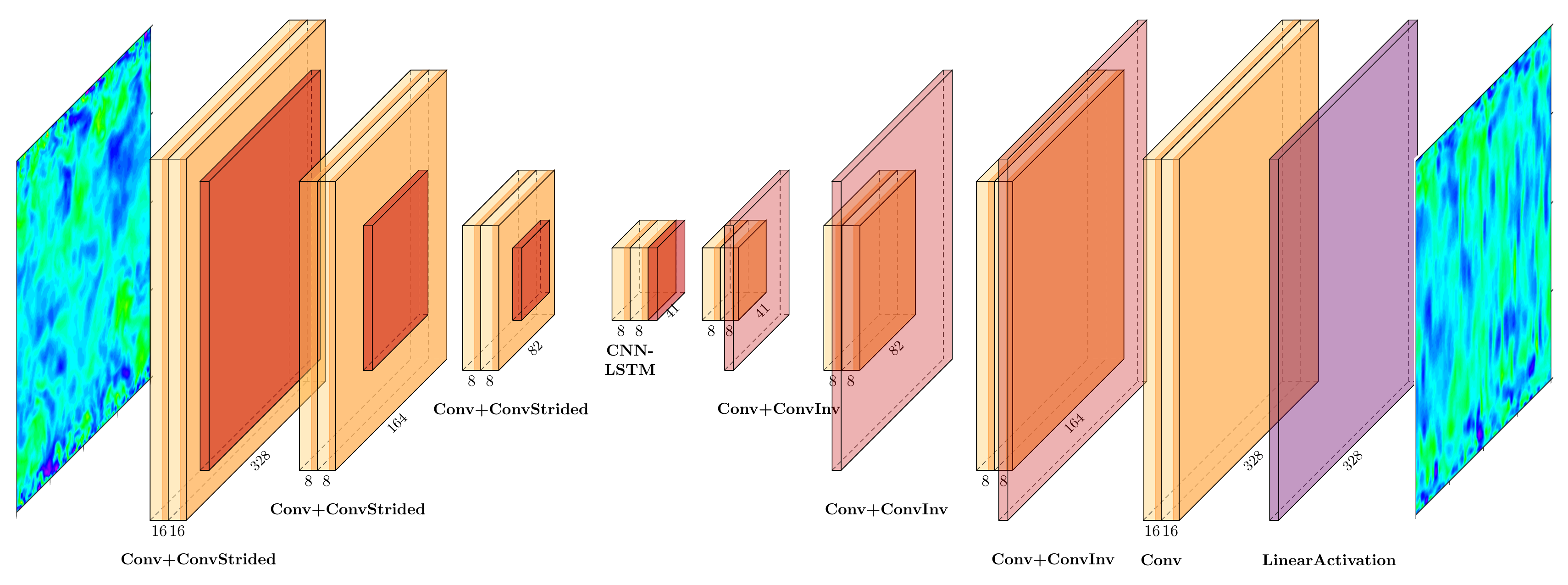} 
         \caption{Deep learning architecture for Case-2 }
    \label{Case2ArchFig}
    \end{figure}

In Case-2, three successive convolutional layers each with $2$ strides were used to reduce the spatial dimensions of input data from  $328 \times 400 $ to $41 \times 50 $, and after the latent-space operation, the same compressed spatial data of  $41 \times 50 $ was then successively upsampled back to its original dimensions of $328 \times 400$ using inverse convolutional operation again with 2 strides. For temporal learning, a convolutional long short-term memory (CNN-LSTM) model was used in the latent-space with the shape of input as well as output as $41 \times 50$. Table \ref{Case2Arch} shows the detailed architecture of this deep learning model. A simplified representation of this model is shown in Figure \ref{Case2ArchFig}. 
For this case, the number of filters again start with $16$, thus the $4$ input channels are mapped to $16$ channels in the CNN making the total dimension of the first layer as $328 \times 400 \times 16$. After successive convolutional layers and by reducing the number of filters to $8$, the latent space dimension becomes $41 \times 50 \times 8$ and same is kept in the CNN-LSTM by using number of strides as 1. Later, the upsampling along with successive convolutions was performed with the same size of filters as that used while dimension reduction. At last, one convolution operation along with a linear activation and $4$ filters produces the output with the same dimensions as that of input. 

\chapter{Analysis of Learned Turbulence}

\section{Visualization of Snapshots}
\begin{figure}[htbp!]
       \centering
        \includegraphics[width=0.99\linewidth, height=5cm]{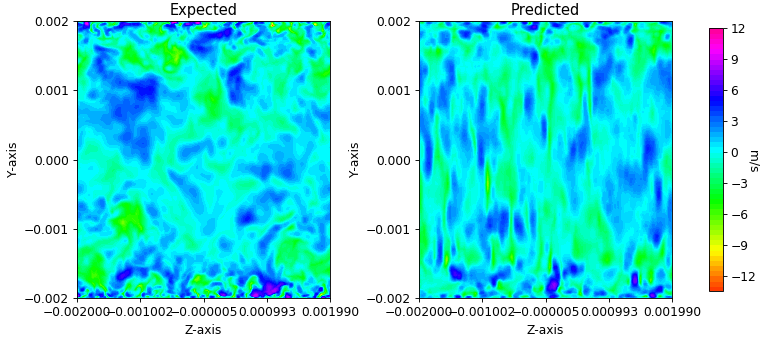} %
        \caption{A snapshot of $u^{\prime}$ at time $0.5N_{iter}$ for Case-1}
        \label{figCase1_visu}
   \end{figure}

Figure \ref{figCase1_visu} shows a snapshot at a certain time instant of the predicted field $u^{\prime}$ from Case-1 compared to the expected field. It was observed from the evolution of snapshots for the \textit{a priori} simulation that the structures in the flow are rapid which could be either due to the reason that network was supplied input data at every 10 iterations of WRLES or because the network as has not learned to produce the proper evolution in time. It is unclear why the network is learning to produce unphysically elongated flow structures. Similar behavior was observed in the the prediction of $v^{\prime}$, $w^{\prime}$, and $p^{\prime}$ fields. The use of CNN-LSTM in the latent-space has significantly improved the results for Case-2 and could be attributed to the way LSTM keeps a record of short history of snapshots. Visualization of the evolution of these predicted fields can be found in Appendix-B. 

\begin{figure}[htbp!]
       \centering
        \includegraphics[width=1.0\linewidth, height=5cm]{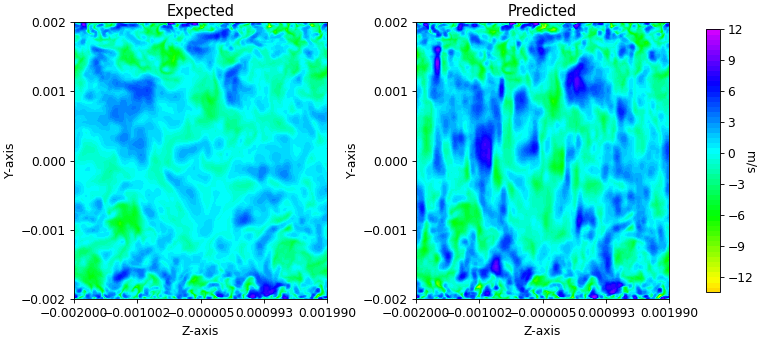} %
        \caption{A snapshot of $u^{\prime}$ at time $0.5N_{iter}$ for Case-2}
        \label{figCase2_visu}
   \end{figure}

Figure \ref{figCase2_visu} shows a snapshot of the predicted field $u^{\prime}$ from Case-2 compared to the expected field. The predicted fields show the fairly good qualitative similarities and distribution of structures compared to WRLES. Similar observations can be drawn about the rapid formation of flow structures. A good agreement was also observed in the prediction of $v^{\prime}$, $w^{\prime}$, and $p^{\prime}$ fields. Visualization of the evolution of these predicted fields can be found in Appendix-B.

\section{Distribution of Fluctuations}
\begin{figure}[htbp!]
       \centering
        \includegraphics[width=1.0\linewidth, height=6cm]{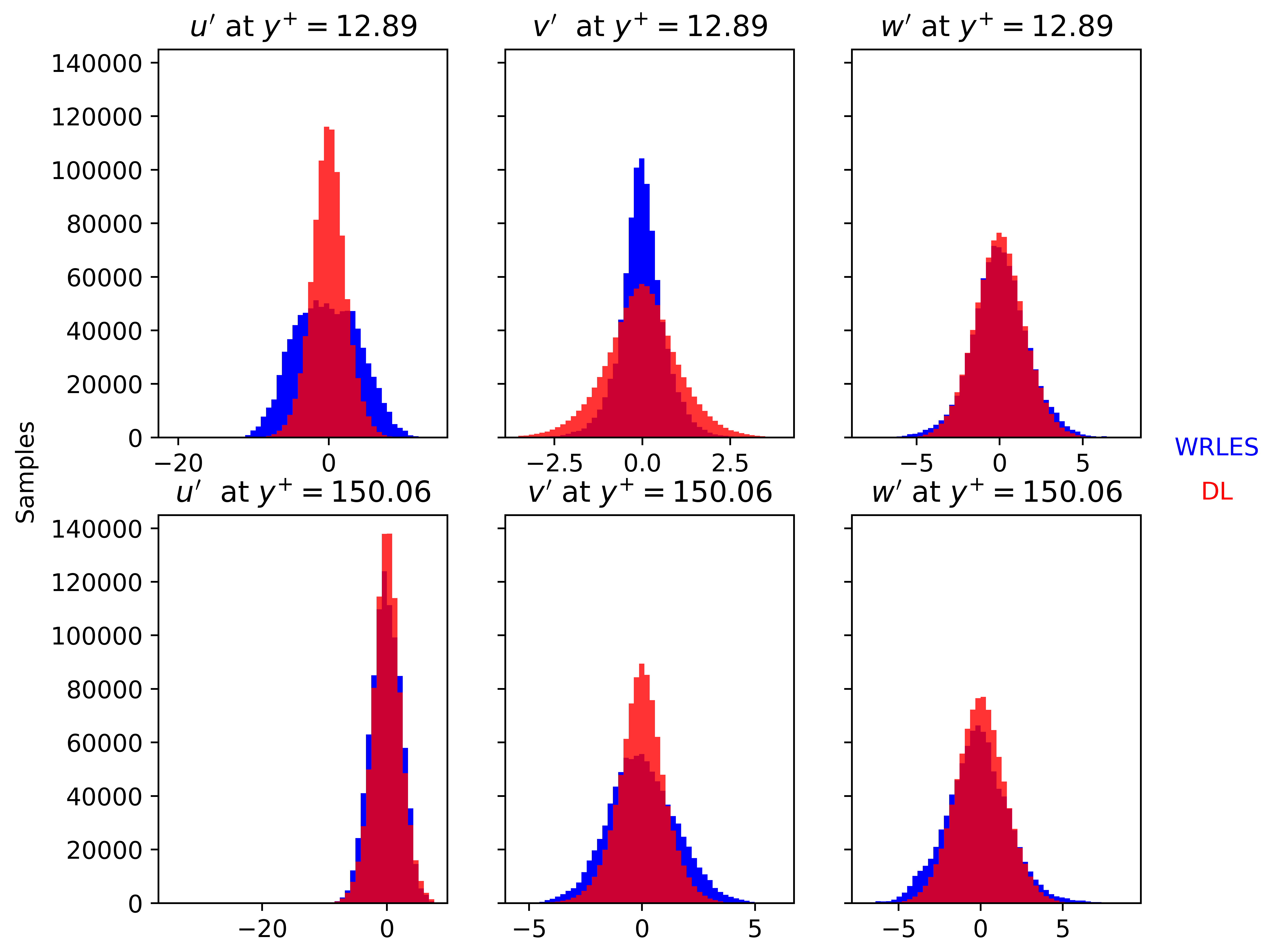} %
        \caption{Comparison of distribution of fluctuations for Case-1. Blue: WRLES distribution, red: learned distribution }
        \label{c1_plot_histData}
   \end{figure}

In order to introduce more quantative assessement, the distribution of velocity components $u^{\prime}$, $v^{\prime}$ , $w^{\prime}$, are investigated. Figure \ref{c1_plot_histData} shows the distribution of learned fluctuations for Case-1 and WRLES fluctuations.
As the current study is that of channel flow, this distribution of fluctuations is being compared at two positions along the y-direction. The top part shows the distribution near the wall at $y^+$=12.89 and the bottom part shows distribution away from the wall at $y^+$=150.06.The $w^{\prime}$ distributions are in good agreement near the wall as well as in the flow. Distribution of $u^{\prime}$ and $v^{\prime}$ are not as good and looks considerably skewed. 

\begin{figure}[htbp!]
       \centering
        \includegraphics[width=0.99\linewidth, height=8cm]{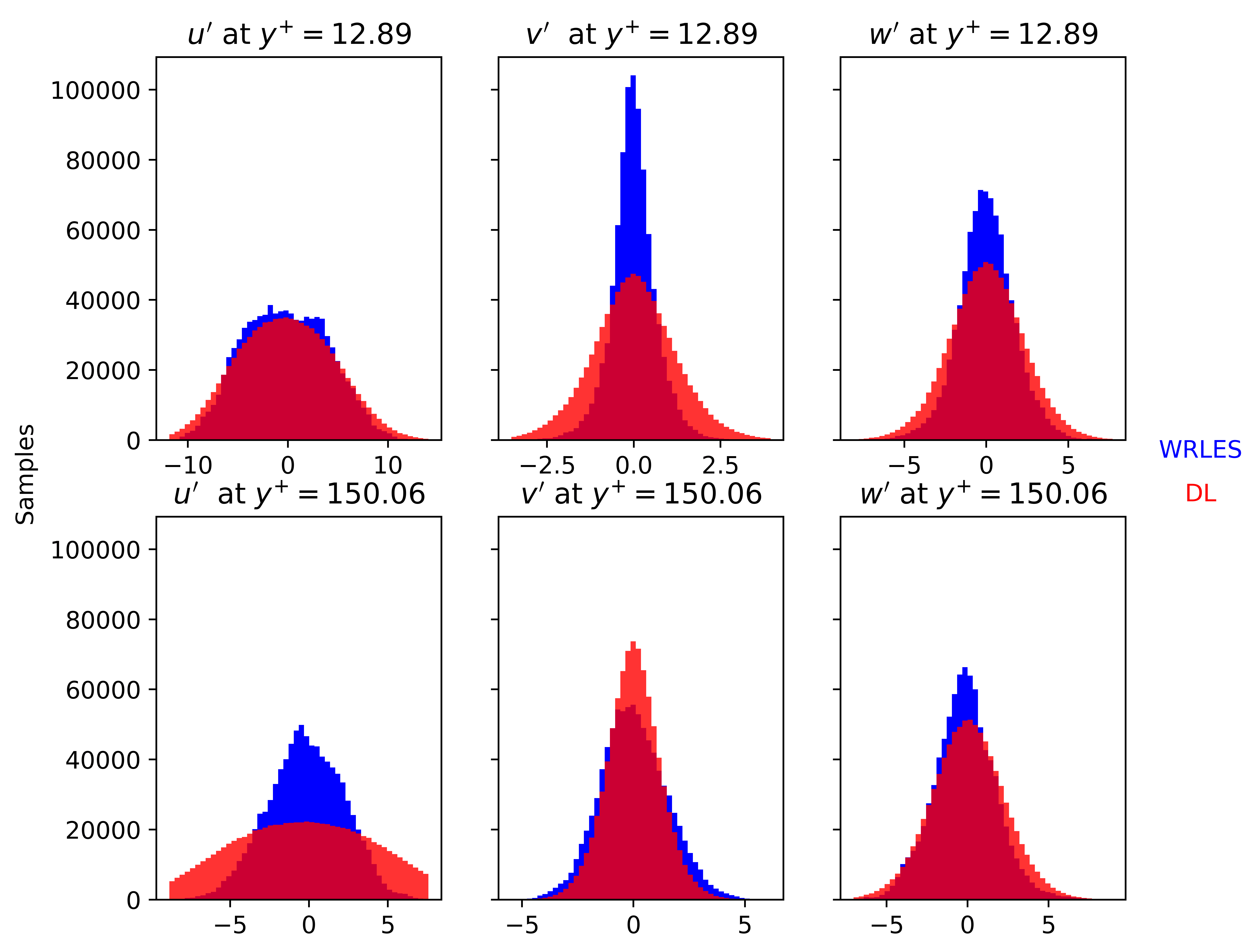} %
        \caption{Comparison of distribution of fluctuations for Case-2. Blue: WRLES distribution, red: learned distribution }
        \label{c2_plot_histData}
   \end{figure}

For Case-2, the distribution of learned and WRLES fluctuations are compared in Figure \ref{c2_plot_histData}. In this case, the neural network has been significantly improved and has learned to produce a good distribution of fluctuations near the wall. Distributions of fluctuations away from the wall are not being re-produced properly. 

\section{RMS Fluctuations}
\begin{figure}[htbp!]
       \centering
        \includegraphics[width=0.99\linewidth, height=7cm]{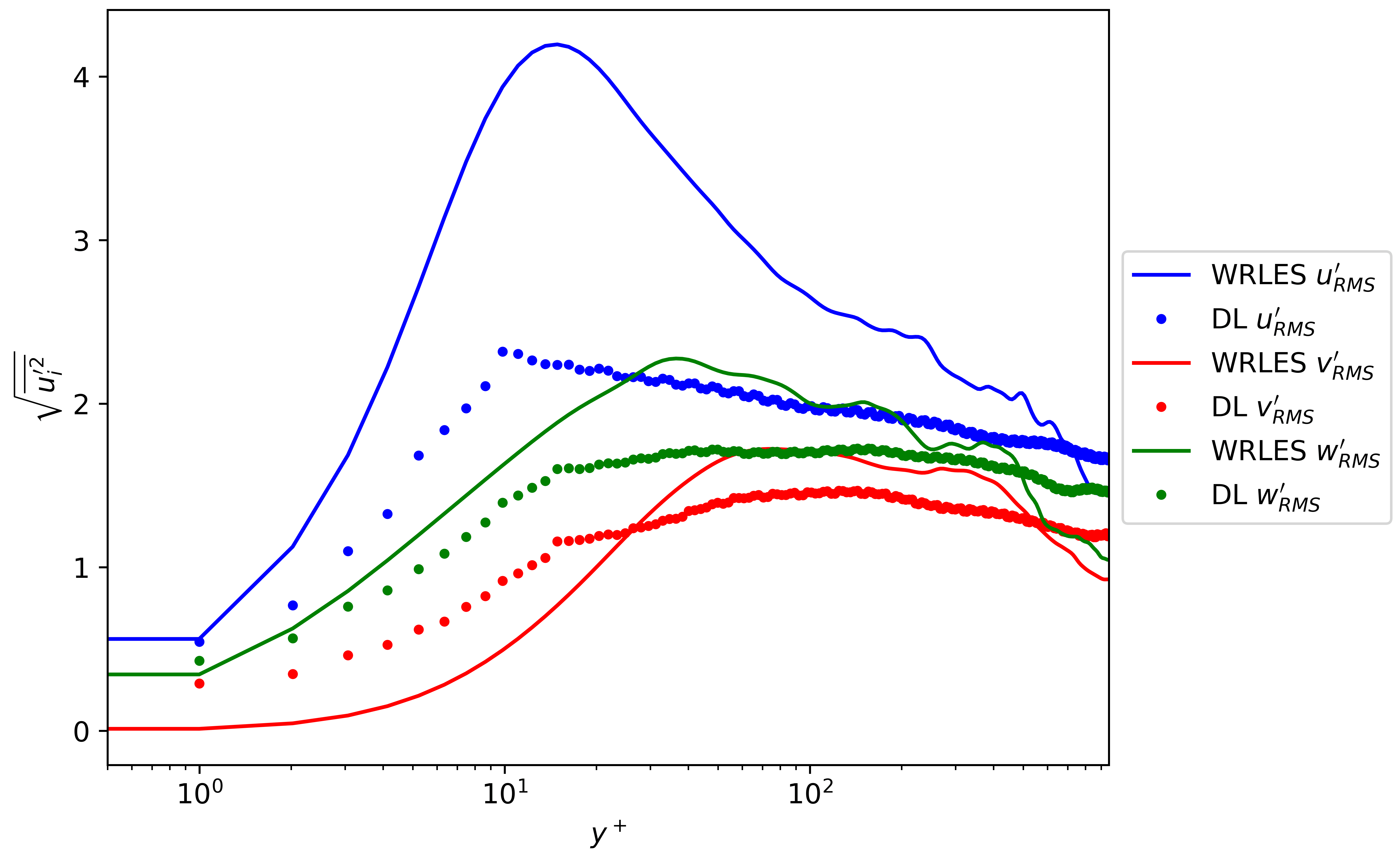} %
        \caption{Comparison of RMS fluctuations $\sqrt{\overline{u_i^{\prime}{}^2}}$ for Case-1}
        \label{c1_plot_RMSFluctuations}
   \end{figure}

\begin{figure}[htbp!]
       \centering
        \includegraphics[width=0.99\linewidth, height=7cm]{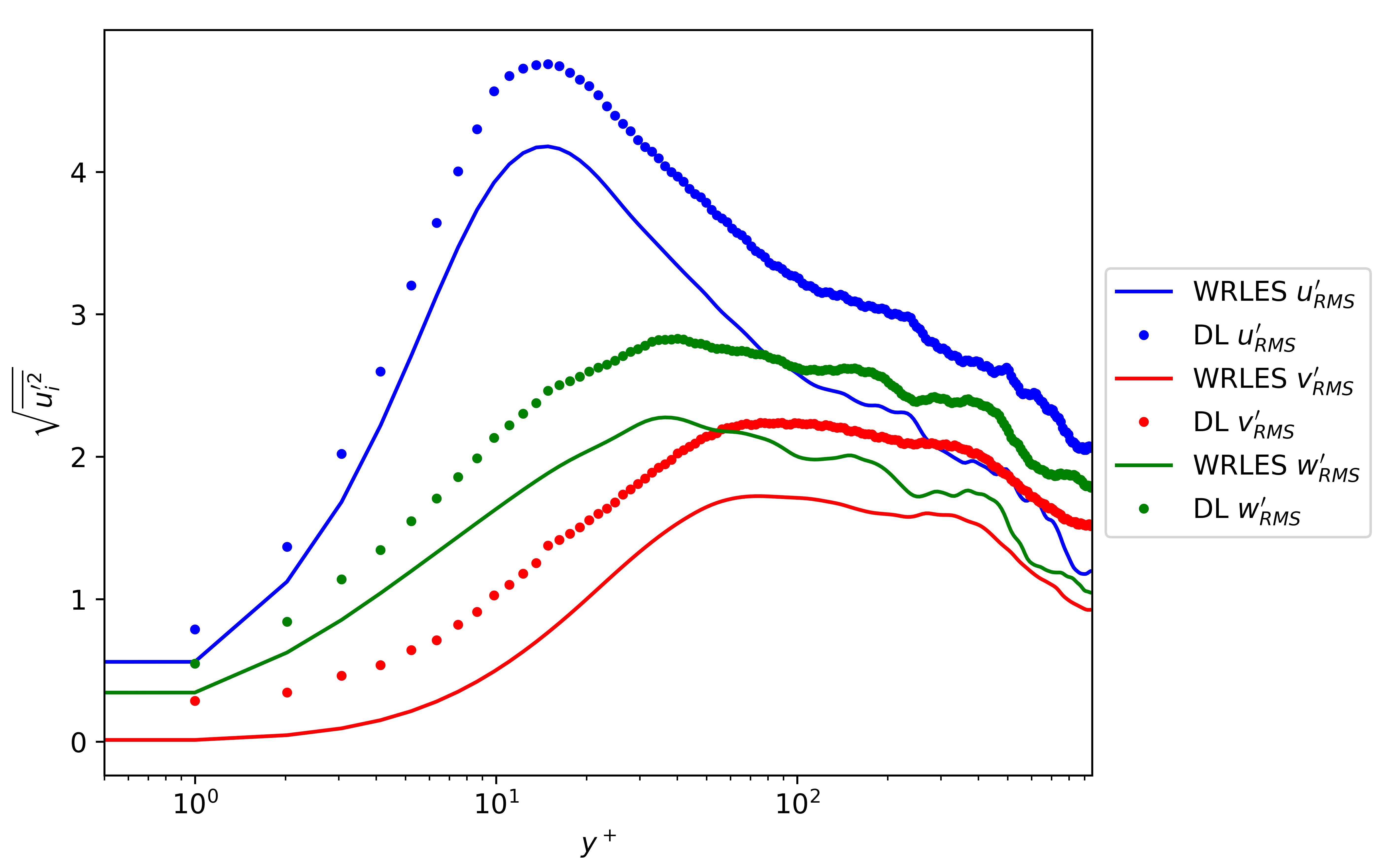} %
        \caption{Comparison of RMS fluctuations $\sqrt{\overline{u_i^{\prime}{}^2}}$  for Case-2}
        \label{c2_plot_RMSFluctuations}
   \end{figure}

To estimate the quality of turbulence produced, the root mean squared (RMS) fluctuations averaged over time are compared for the three velocity components $u^{\prime}$, $v^{\prime}$, $w^{\prime}$ in Figure \ref{c1_plot_RMSFluctuations} and Figure \ref{c2_plot_RMSFluctuations}. 
For Case-1, the neural network has sufficiently learned the shape of these fluctuations but failed to learn their magnitudes. 
Case-2, shows a better agreement with WRLES, but the magnitude is still-off by a certain extent away from the wall. 

\section{Reynolds Stresses}
\begin{figure}[htbp!]
       \centering
        \includegraphics[width=1.00\linewidth, height=7cm]{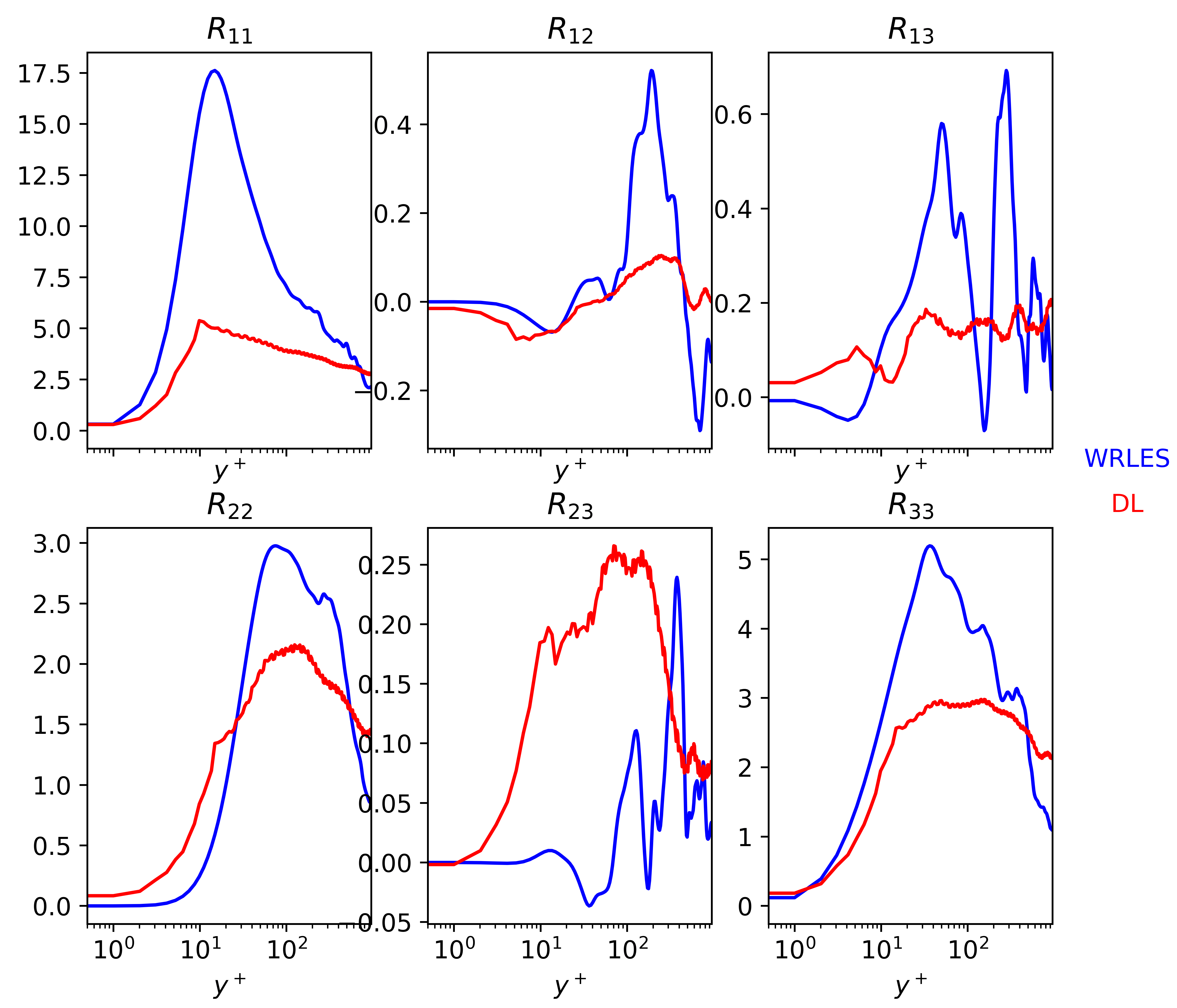} %
        \caption{Comparison of Reynolds Stresses $R_{ij}$ for Case-1}
        \label{c1_plot_yPlus_Rij}
   \end{figure}

\begin{figure}[htbp!]
       \centering
        \includegraphics[width=1.00\linewidth, height=7cm]{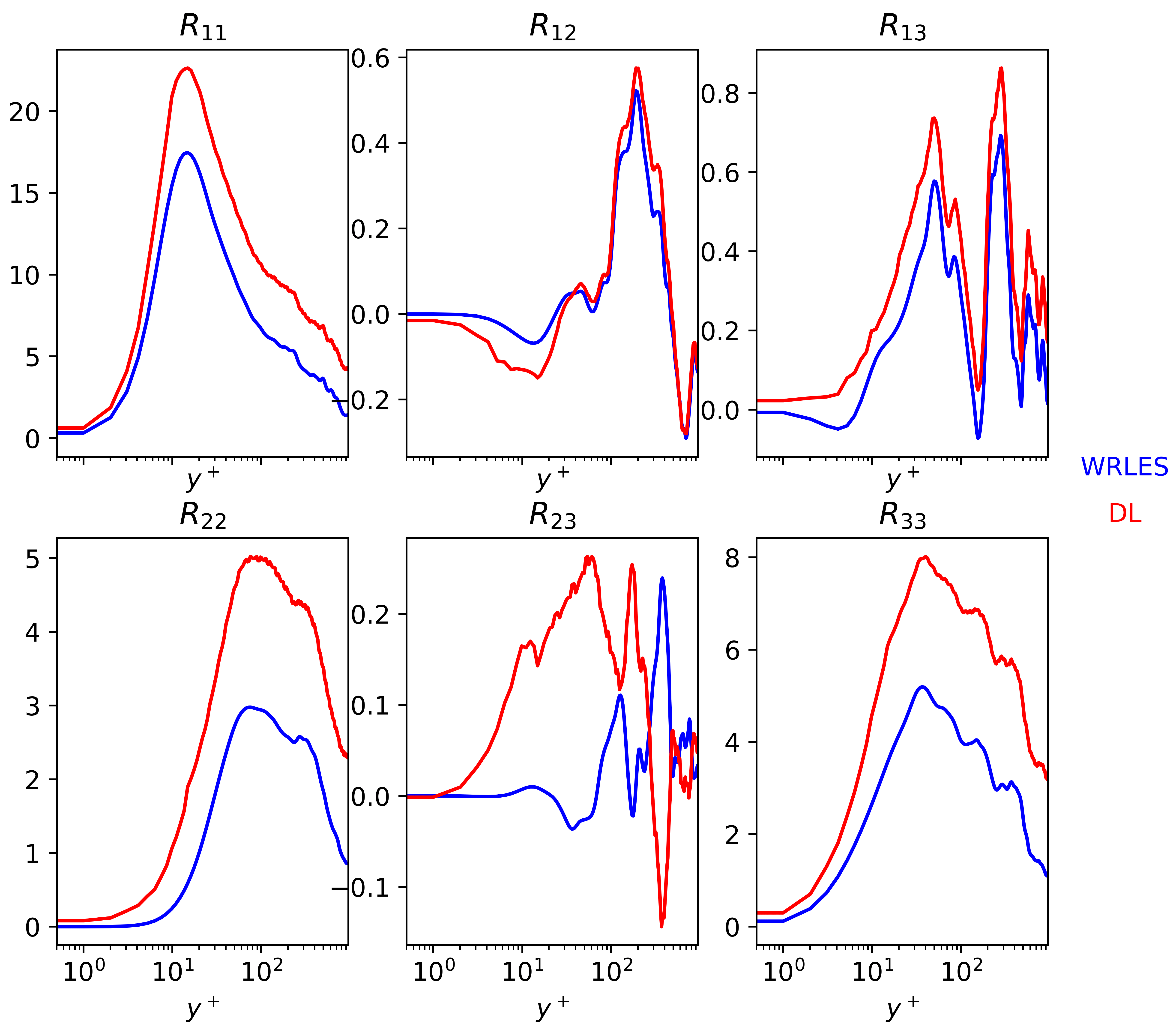} %
        \caption{Comparison of Reynolds Stresses $R_{ij}$ for Case-2}
        \label{c2_plot_yPlus_Rij}
   \end{figure}

Components of Reynolds stresses give a good insight into the physics occuring near the wall. Reynolds stresses, $R_{11}$,$R_{12}$,$R_{13}$,$R_{22}$,$R_{23}$,$R_{33}$ are computed for the neural network \textit{a priori} simulation Case-1 and compared with WRLES data in Figure \ref{c1_plot_yPlus_Rij}. The model in this case has learnt to produce a good shape but with wrong magnitudes near the wall. The order of magnitudes of Reynolds stress components, $R_{11}$ > $R_{33}$ > $R_{22}$ > $R_{13}$ > $R_{12}$ > $R_{23}$, is also preserved by the model which is also evident from the RMS fluctuations. 
For Case-2, as shown in Figure \ref{c2_plot_yPlus_Rij}, the shape and magnitude for all Reynolds stress components except $R_{23}$ are in good agreement with the WRLES data. 

\section{Velocity Correlations}
\begin{figure}[htbp!]
       \centering
        \includegraphics[width=0.48\linewidth, height=6cm]{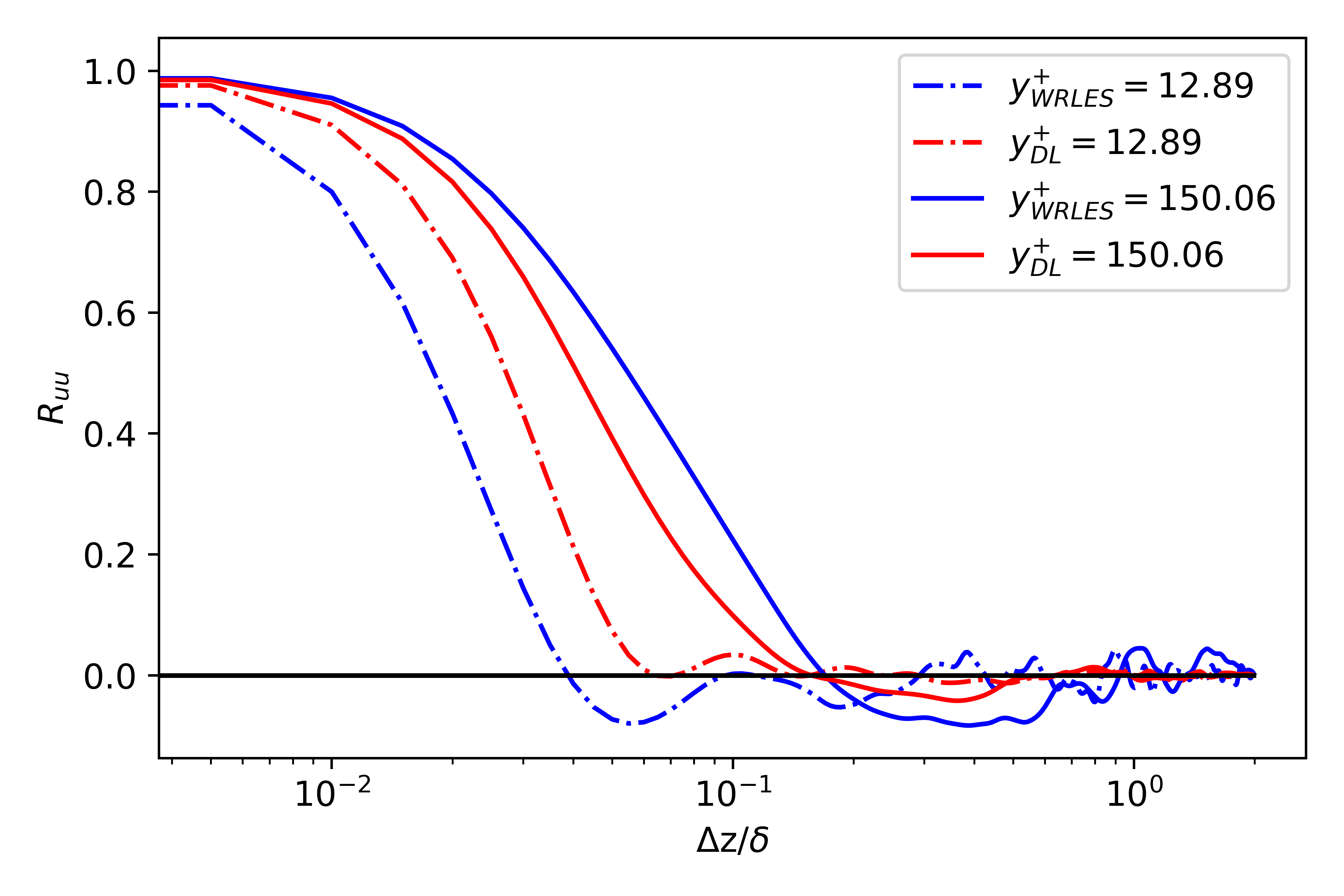} 
        \includegraphics[width=0.48\linewidth, height=6cm]{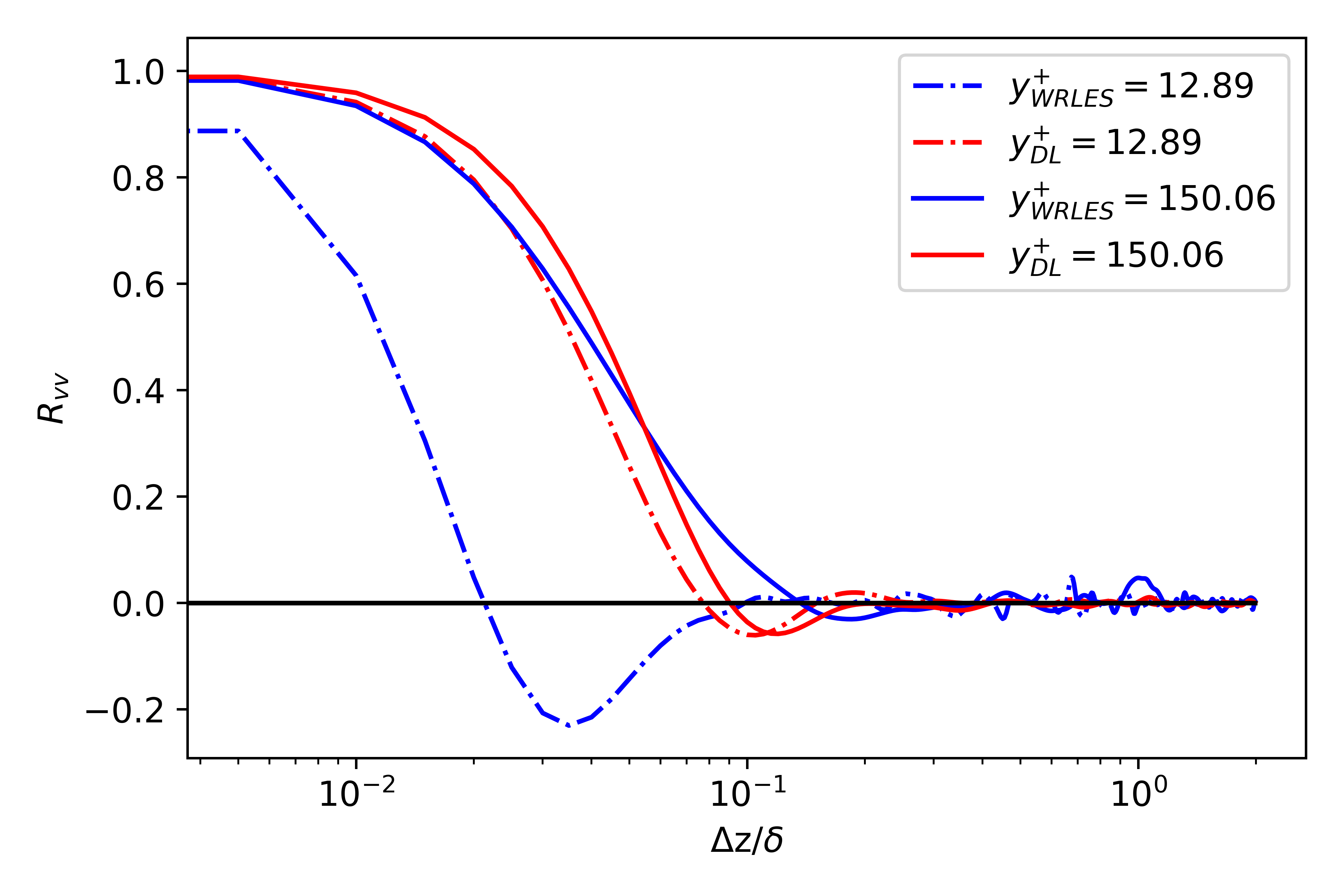} 
        \caption{$R_{uu}$ (left) and $R_{vv}$ (right) correlations for Case-1}
        \label{c1_plot_autoCorr_Rvvuu_tAvg}
   \end{figure}

Velocity correlations are an effective tool for giving insight into the nature of turbulence developed. As the data in the current study is in the $y-z$ plane, span-wise velocity correlations at a certain $y^+$ position with ensemble average in time were computed as:
\begin{equation}
R_{ij} (y,\Delta z) = \frac{< u_i(z,t) u_j(z+\Delta z,t)  >}{<u_i^2 (z,t)>^{\frac{1}{2}} <u_j^2 (z+\Delta z, t)>^{\frac{1}{2}} }  
\end{equation}

such that $R_{uu}$ is the correlation of $u^{\prime}$ velocity component, $R_{vv}$ is the correlation of $v^{\prime}$ velocity component and both are averaged for all the time instants which correspond to approximately $N_{iter}$ iterations for the respective cases. For Case-1, both the $R_{uu}$ and $R_{vv}$ correlations are computed at $y^+$ = 12.89 and at $y^+$=150.06 which are shown in Figure \ref{c1_plot_autoCorr_Rvvuu_tAvg}(left) and Figure \ref{c1_plot_autoCorr_Rvvuu_tAvg}(right). The dashed lines and solid lines show the correlations near the wall and away from the wall respectively, whereas the red lines show results obtained from the deep learned model and blue lines show the results obtained from WRLES. It can be observed that the correlations obtained from the deep learned model are not in complete agreement with the correlations from WRLES data. The closeness of correlation curves obtained from deep learned model shows that the integral length scales near the wall and away from the walls both have almost the same magnitude. This is also evident from the visualization of snapshots Figure \ref{figCase1_visu}) which show the same size of structures produced everywhere. This could be an effect of the MLP being used in the latent space which has no memory of snapshots in time, hence producing the same size of structures again and again.

\begin{figure}[htbp!]
       \centering
        \includegraphics[width=0.48\linewidth, height=6cm]{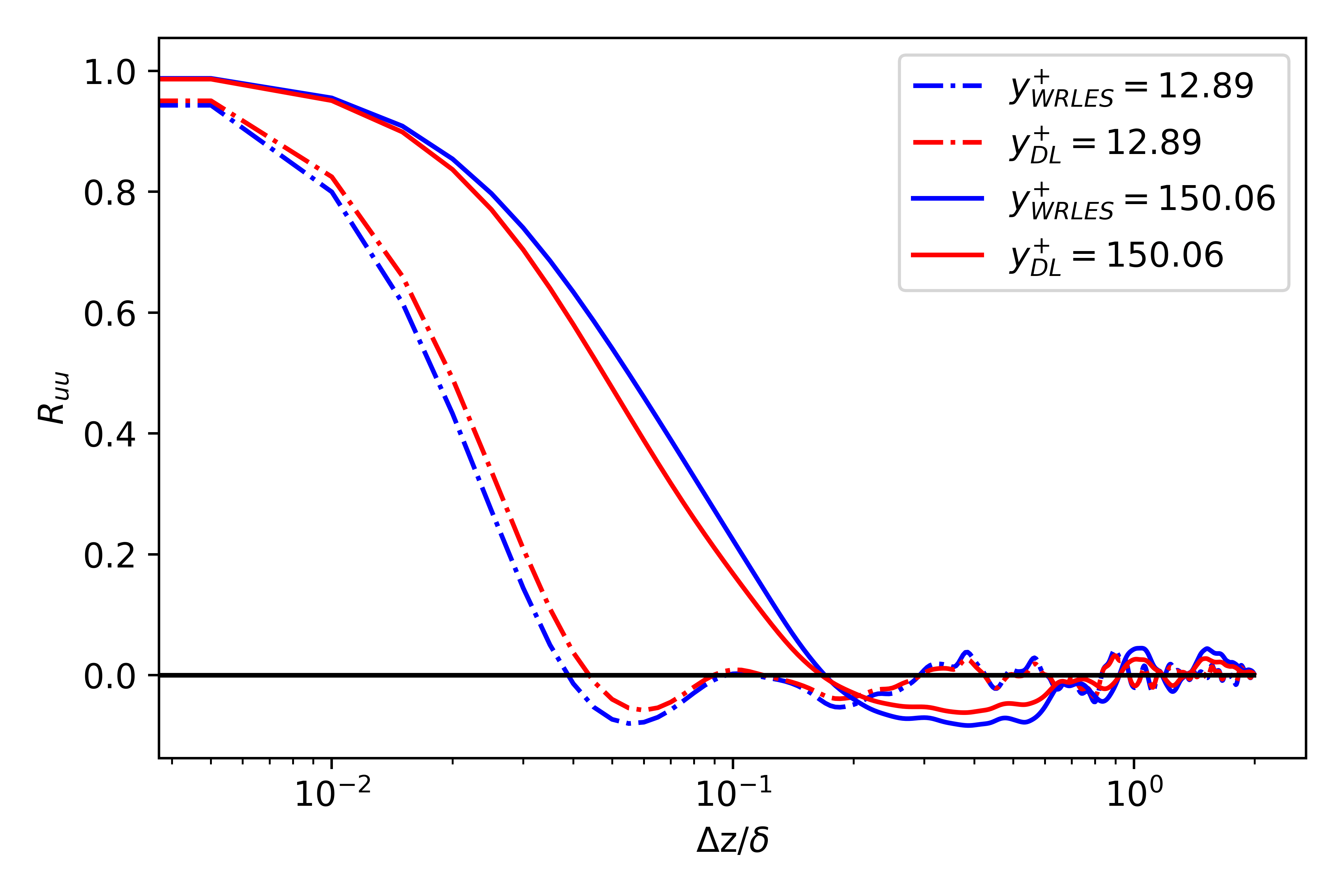} 
        \includegraphics[width=0.48\linewidth, height=6cm]{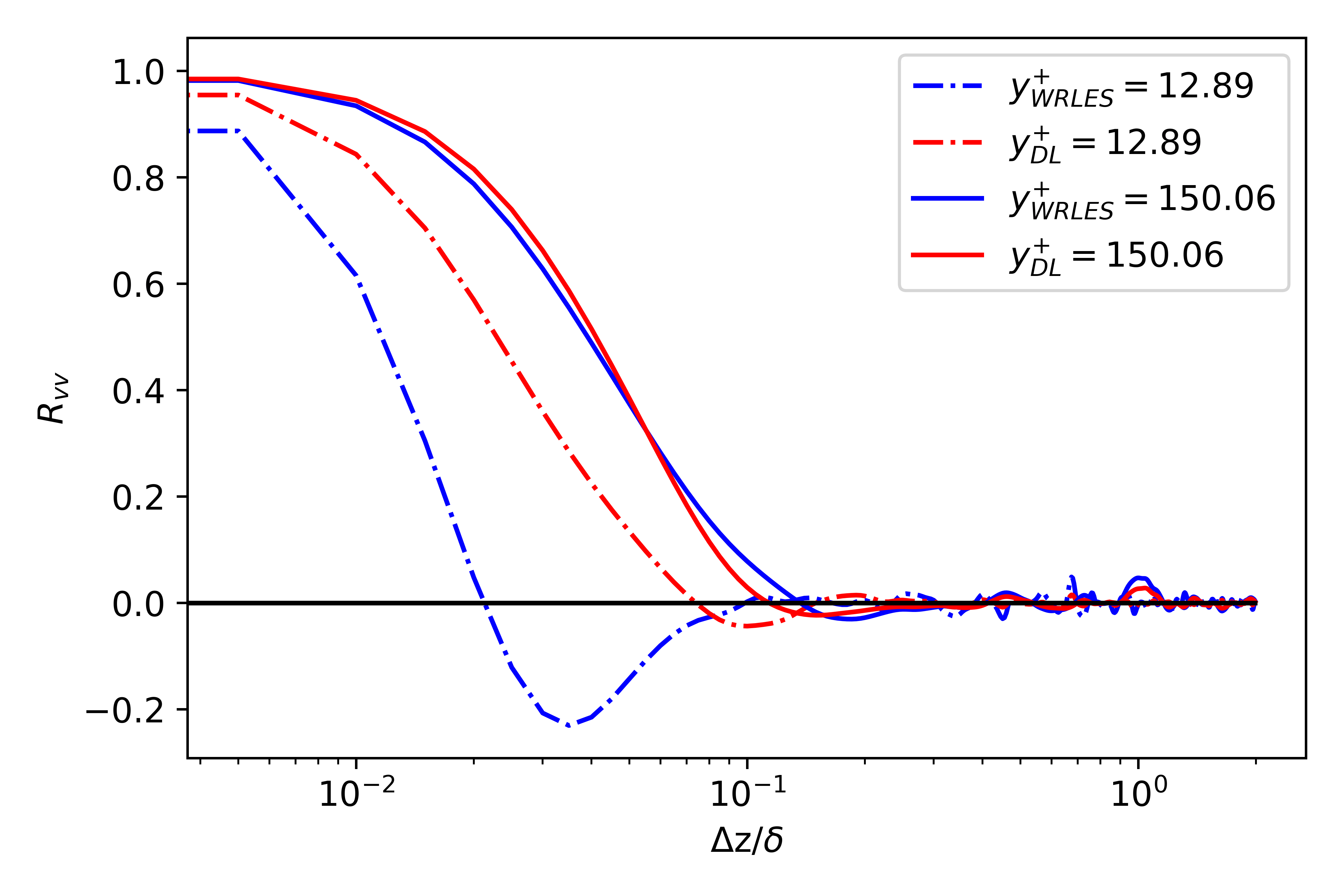} 
        \caption{$R_{uu}$ (left) and $R_{vv}$ (right) correlations for Case-2}
        \label{c2_plot_autoCorr_Ruuvv_tAvg}

   \end{figure}

Similarly, the $R_{uu}$ and $R_{vv}$ correlations are computed again at the same near-wall and away from the wall locations i.e. $y^+$ = 12.89 and $y^+$=150.06 for Case-2 as shown in Figure \ref{c2_plot_autoCorr_Ruuvv_tAvg}(left) and Figure \ref{c2_plot_autoCorr_Ruuvv_tAvg}(right). It can be noted that the correlations from the learned model, in this case, are significantly better and match the WRLES data. Correlations at both near-wall and away from the wall locations are in a good agreement with the WRLES data. This suggests that the model has learned to estimate the evolving nature of flow structures better than Case-1 with the help of a CNN-LSTM in the latent space.

\section{Turbulent Kinetic Energy Spectra }

\begin{figure}[htbp!]
       \centering
        \includegraphics[width=0.48\linewidth, height=6cm]{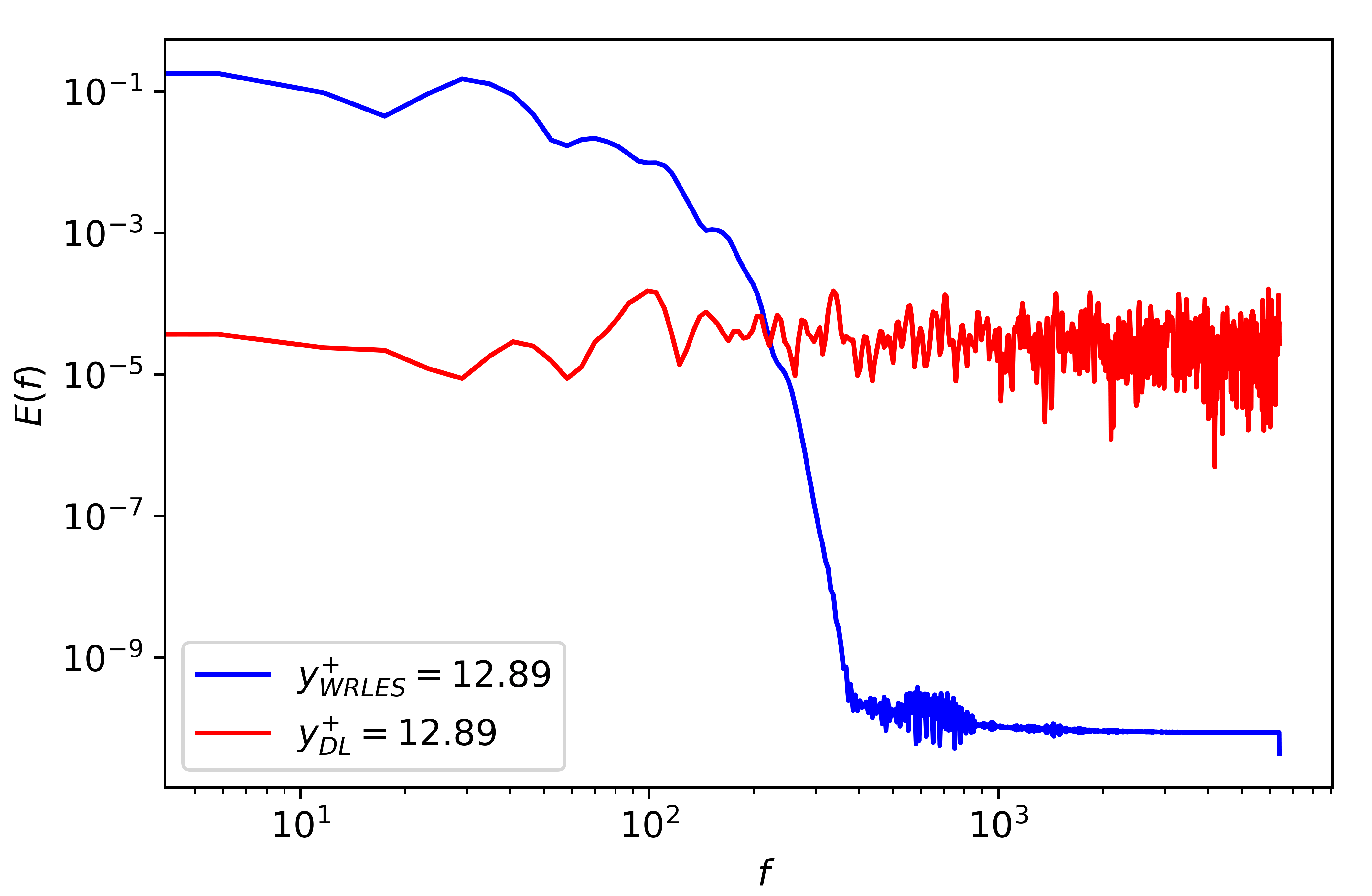} 
        \includegraphics[width=0.48\linewidth, height=6cm]{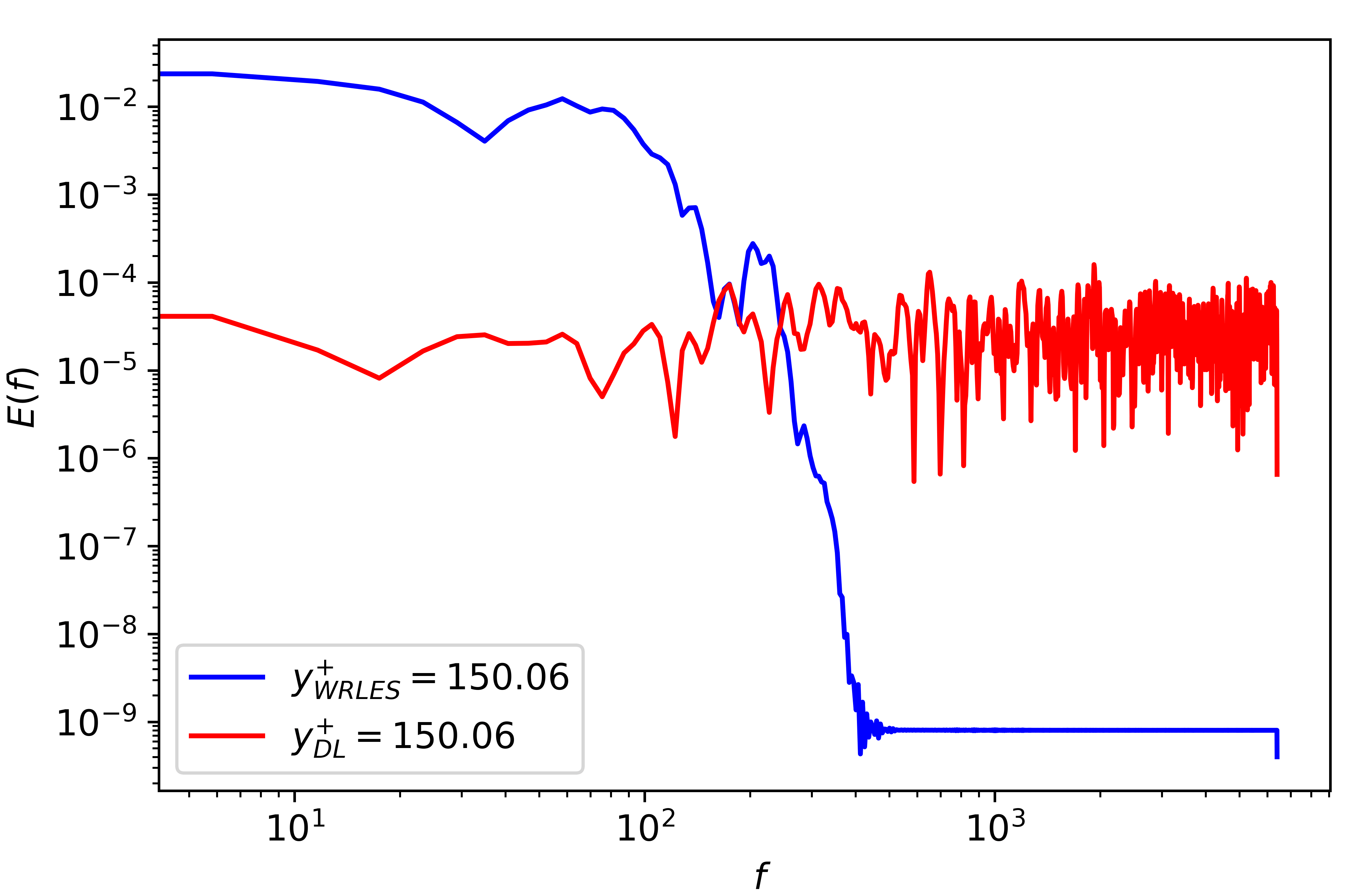} 
        \caption{Comparison of spanwise turbulent kinetic energy spectrum at two wall-normal positions for Case-1.  Spectrum at $y^+$=12.89 (left)  and spectrum $y^+$=150.06 (right)}
        \label{c1_plot_TKEspectrum_zAvg}

   \end{figure}

Spectral analysis of turbulent kinetic energy is an essential tool in accessing the multi-scale nature of turbulence. The turbulence generated from the deep learning model's \textit{a priori} simulation should ideally have the same spectral contents as of real turbulent flows so as to sustain the turbulence fluctuations for longer durations. To estimate this, spanwise turbulent kinetic energy spectrum ($E(f)$) is computed at two different $y^+$ positions. The spectrum obtained is ensembled averaged over all the snapshots.  %
Figure \ref{c1_plot_TKEspectrum_zAvg}(left) and Figure\ref{c1_plot_TKEspectrum_zAvg}(right) shows the plot of energy spectrum for Case-1. The plot to the left is the spectrum at $y^+$=12.89 and the plot to the right shows the spectrum at $y^+$=150.06; the red line shows the spectrum computed from the deep learning model whereas blue line shows the WRLEs spectrum. For this case, it can be observed that the spectral energy content in the turbulence produced from deep learned model is not consistent with the WRLES results both near the wall and away from the wall. The same level of energy is being accumulated across all the scales which is again inconsistent with the WRLES spectrum. 

\begin{figure}[htbp!]
       \centering
        \includegraphics[width=0.48\linewidth, height=6cm]{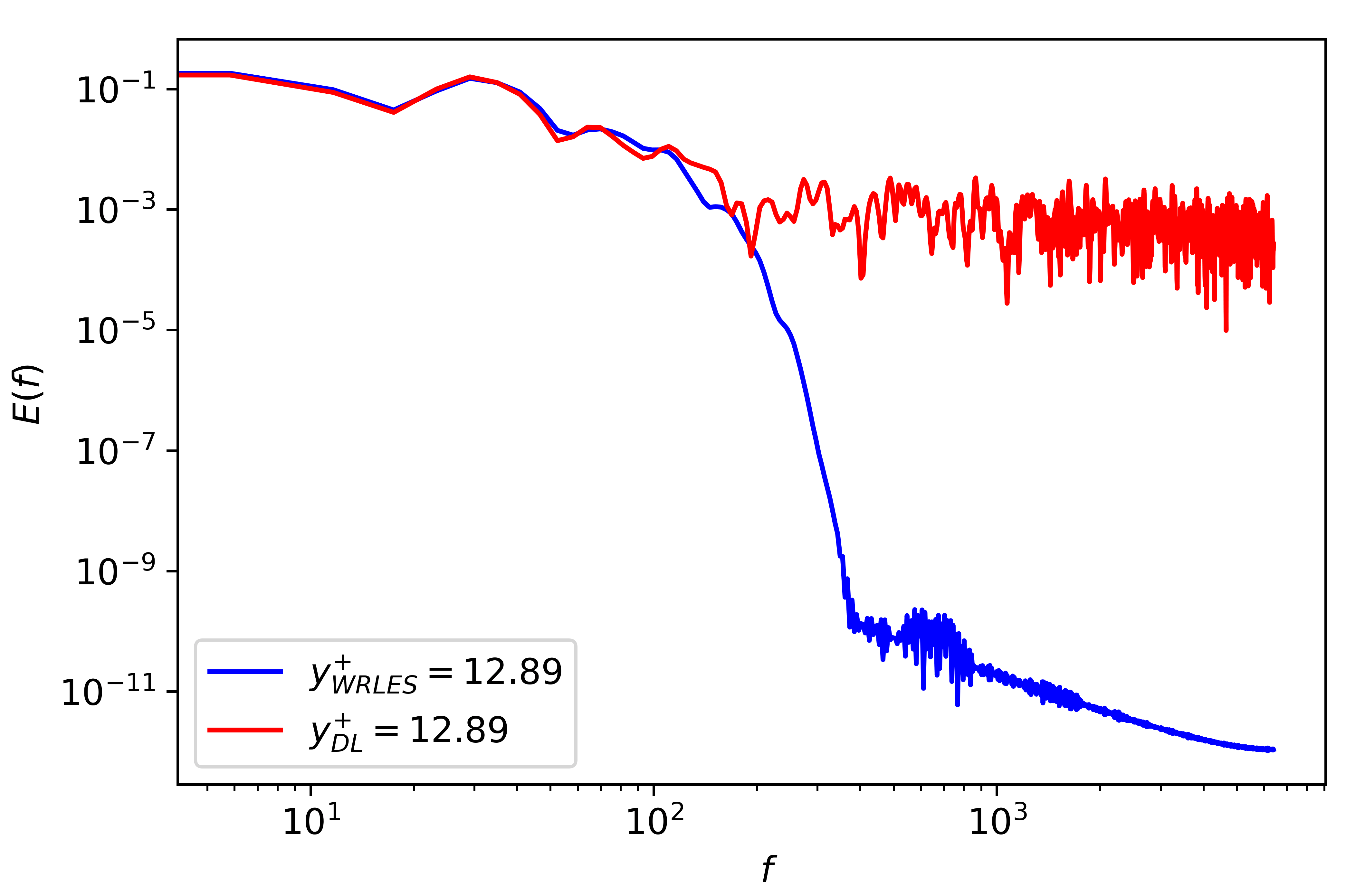} 
        \includegraphics[width=0.48\linewidth, height=6cm]{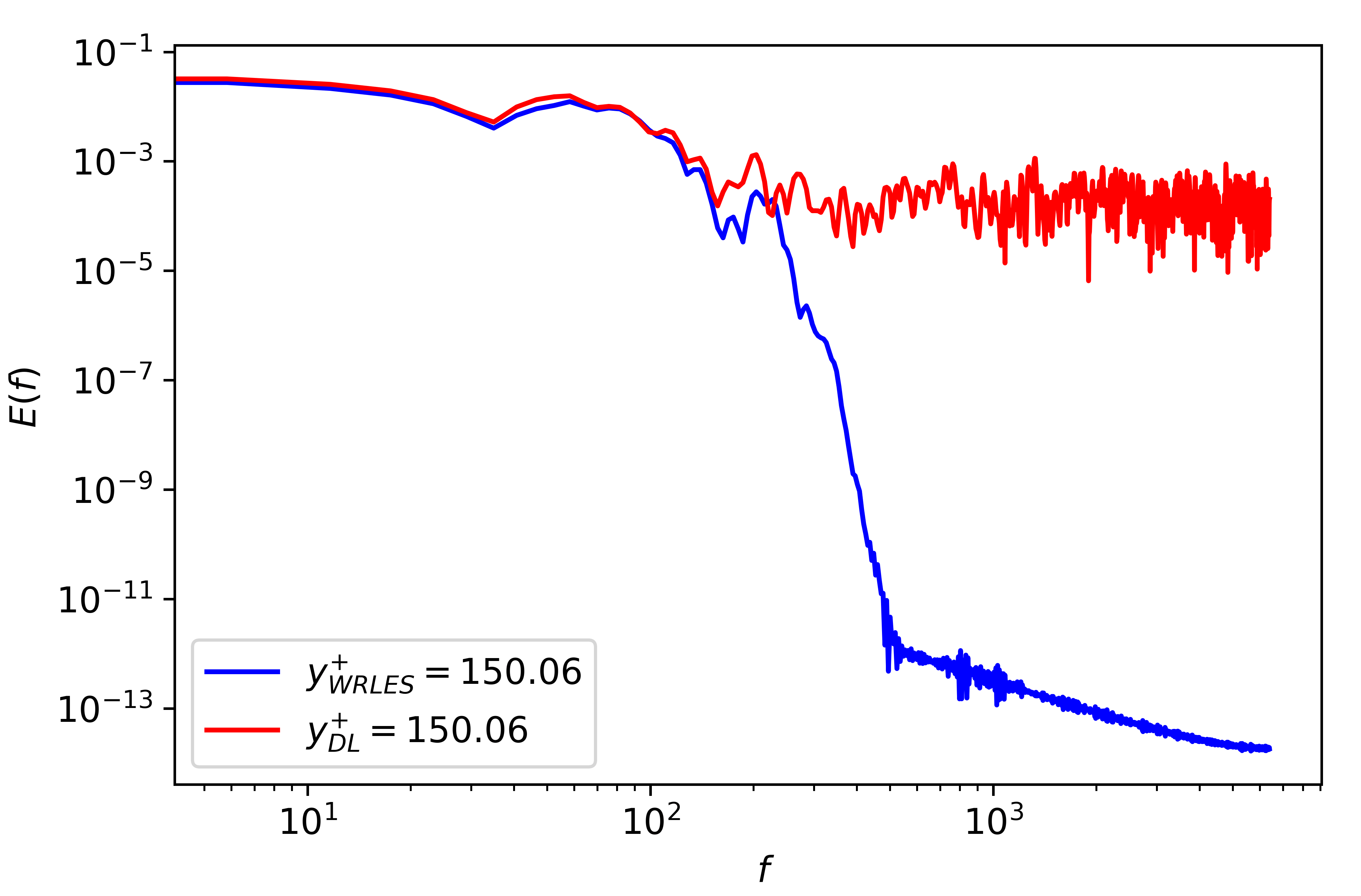} 
        \caption{Comparison of spanwise turbulent kinetic energy spectrum at two wall-normal positions for Case-2. Spectrum at $y^+$=12.89 (left)  and spectrum $y^+$=150.06 (right)}
        \label{c2_plot_TKEspectrum_zAvg}

   \end{figure}

For Case-2, as plotted in Figure \ref{c2_plot_TKEspectrum_zAvg}(left) and Figure \ref{c2_plot_TKEspectrum_zAvg}(right) , it is can be noted that the spectrum computed from the deep learning model is considerably better as compared to the WRLES spectrum. Both near wall and away from the wall spectrum show good agreement with the WLRES spectrum at the large scales. The model fails to capture the dissipation of energy to the small scales which means that the model does not take into account the most cut-off frequencies. It appears as if the same amount of energy gets accumulated across all the intermediate and small scales. 

\section{Learned Pressure Fluctuations}
\begin{figure}[htbp!]
       \centering
        \includegraphics[width=1.00\linewidth, height=10cm]{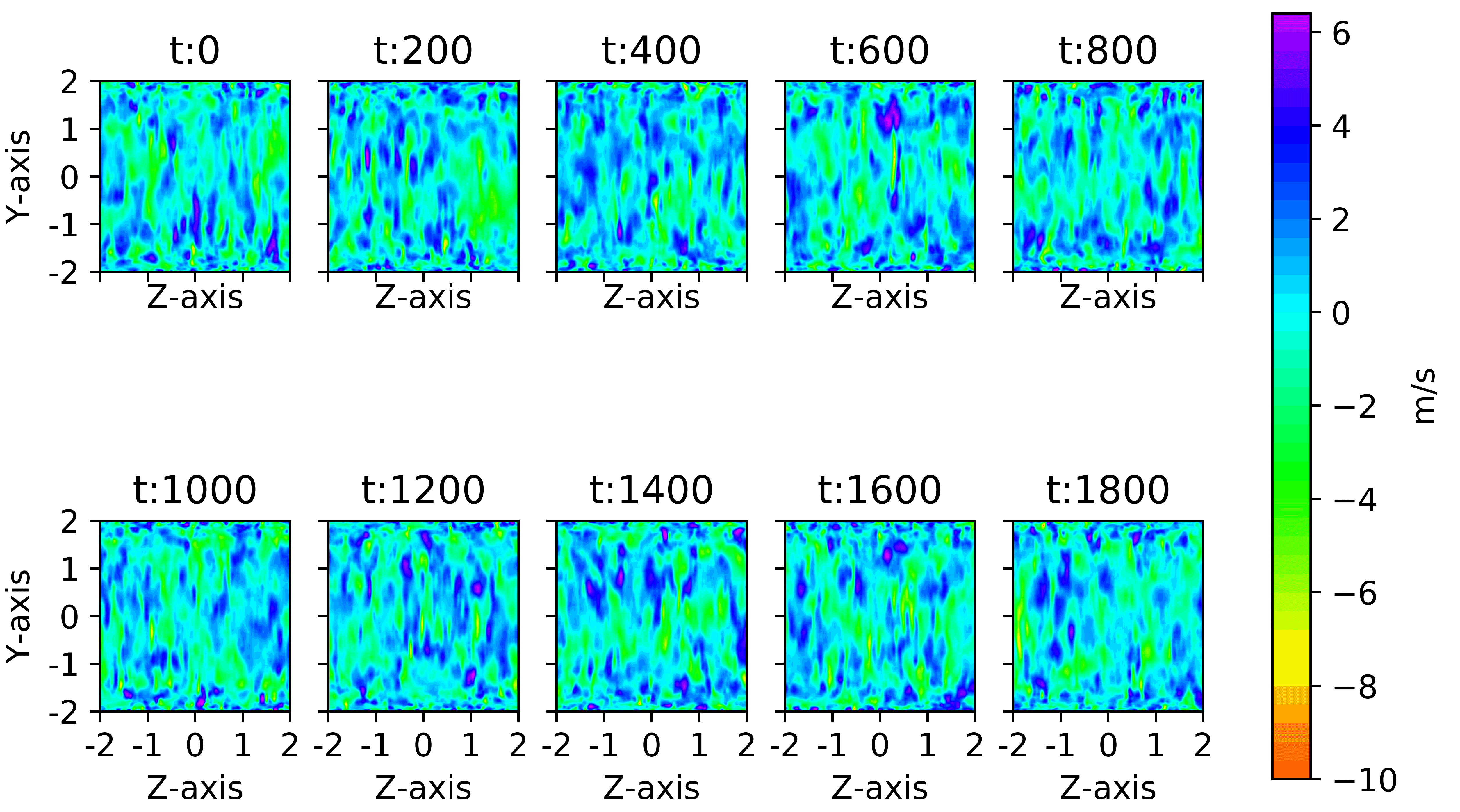}
        \caption{Learned pressure fluctuations for Case-1}
        \label{c1_plot_predicted_aPriori_ch3}
   \end{figure}

\begin{figure}[htbp!]
       \centering
        \includegraphics[width=1.00\linewidth, height=10cm]{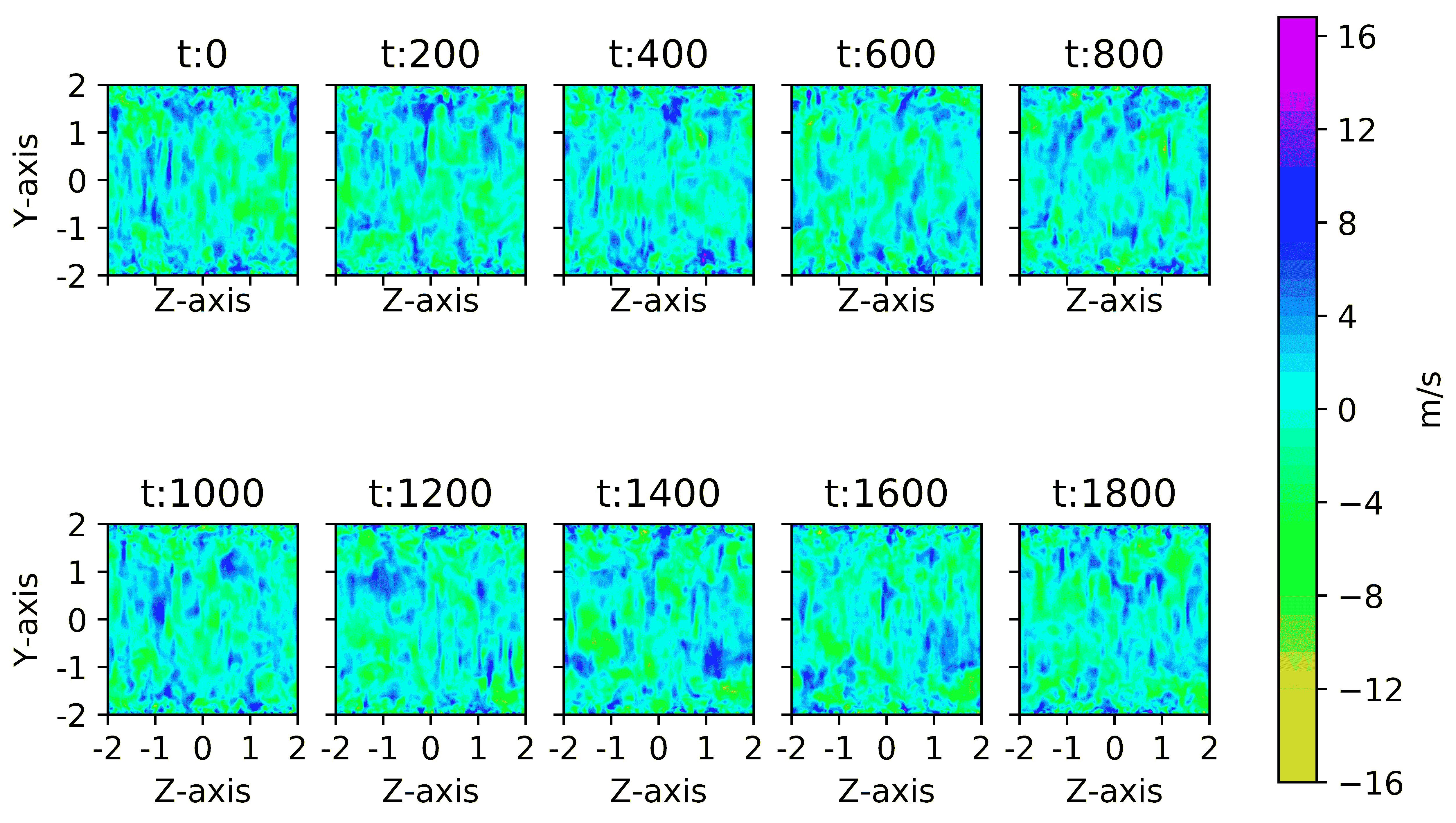}  
        \caption{Learned pressure fluctuations for Case-2}
        \label{c2_plot_predicted_aPriori_ch3}
   \end{figure}

The learned pressure fluctuations are qualitatively compared in Figure \ref{c1_plot_predicted_aPriori_ch3}, which shows an evolution of pressure fluctuations for Case-1. It can be noted that the magnitude of learned pressure fluctuations appears to be consistent with that of WRLES data. Moreover, for Case-2 as shown in Figure \ref{c2_plot_predicted_aPriori_ch3}, the pressure fluctuations obtained from the deep learning model appears to be more physical in terms of magnitude as well as the shape of visual structures. A detailed quantitative analysis is needed to describe the pressure fluctuations obtained from the deep learning models. 

\chapter{Conclusions}
The present work proposes an inflow turbulence generation strategy using deep learning methods. This was achieved with the help of an autoencoder architecture with two different types of operational layers in the latent-space: a fully connected multi-layer perceptron and convolutional long short-term memory layers. A wall-resolved large eddy simulation of a turbulent channel flow at $Re_\tau$=950 was performed to create a large database of instantaneous snapshots of turbulent flow-fields used to train neural networks.
The training was performed with sequences of instantaneous snapshots so as to learn a snapshot at the next time instant and the accuracy of learning was compared by evaluating the mean-squared error. A further investigation is required on the way of comparing the accuracy of learning the turbulent flow-fields. For the present work, velocity and pressure fluctuations were used for training to produce the same parameters at the next time instant. Future work could explore other combinations of flow parameters to produce the inflow turbulence with quantities of interest. Within the autoencoders, different types of methods like average pooling and strided convolutions were tested for the reduction of spatial dimensions. More investigation is needed to ascertain the use of one method over another in terms of capturing the turbulent scales of interest. For the convolutional neural networks, though the physical boundary conditions were implemented in the form of symmetric as well as periodic paddings, an explicit implementation of these boundary conditions needs to be tested. 
\textit{A priori} simulations were performed with the trained deep learning models to check the accuracy of turbulence statistics produced. It was found that the use of convolutional long short-term memory layers in the latent space improved the quality of statistics, although issues related to stability for longer times were observed. Though instantaneous snapshots of the target flow are required for training, these \textit{a priori} simulations suggest that the deep learning methods for generating inflow turbulence are one of the possible alternatives to existing methods. 
A follow-up to the present study could be to introduce the trained neural network into the LES solver to compute the turbulent inflow generation. This would yield the means for \textit{a posteriori} validation of the resulting turbulent flow.

\printbibliography[heading=bibintoc]

\appendix

\chapter{Detailed Structure of Deep Learning Architectures}

\subsubsection{MLP-Latent Space}
\begin{table}[htpb!]
\begin{center}
\caption{{Detailed structure of deep learning architecture for Case-1}}
\def~{\hphantom{0}}
\begin{tabular}{ccc}
    \hline \hline 
    Layer & Output Data Shape & Activation Function \\ \hline
    Input &(328,400,4) & - \\
    1st Conv2D&(328,400,16)& ReLU\\
    2nd Conv2D&(328,400,16)& ReLU \\
    1st AveragePooling 2D&(164,200,16)& - \\
    3rd Conv2D&(164,200,8)& ReLU \\
    4th Conv2D&(164,200,8)& ReLU \\
    2nd AveragePooling 2D&(82,100,8)& - \\
    5th Conv2D&(82,100,8)& ReLU \\
    6th Conv2D&(82,100,8)& ReLU \\
    3rd AveragePooling 2D&(41,50,8)& - \\
    1st Reshape&(1,16400)&-\\
    1st {MLP}&(16400)&ReLU\\
    2nd {MLP} &(16400)&ReLU\\
    2nd Reshape&(41,50,8)&-\\
    7th Conv2D&(41,50,8)& ReLU \\
    8th Conv2D&(41,50,8)& ReLU \\
    1st Upsampling 2D&(82,100,8)& - \\
    9th Conv2D&(82,100,8)& ReLU \\
    10th Conv2D&(82,100,8)& ReLU \\
    2nd Upsampling 2D&(164,200,8)& - \\
    11th Conv2D&(164,200,8)& ReLU \\
    12th Conv2D&(164,200,8)& ReLU \\
    3rd Upsampling 2D&(328,400,16)& - \\
    13th Conv2D&(328,400,16)& ReLU \\
    Output/14th Conv2D&(328,400,4)& Linear \\
    \hline \hline
\end{tabular}
  \label{Case1Arch}
\end{center}
\end{table}
\newpage
\subsubsection{CNN-LSTM Architecture}
\begin{table}[htpb!]
\begin{center}
\caption{{Detailed structure of deep learning architecture for Case-2}}
\def~{\hphantom{0}}
\begin{tabular}{ccc}
    \hline \hline 
    Layer & Output Data Shape & Activation Function \\ \hline
    Input &(328,400,4) & - \\
    1st Conv2D&(328,400,16)& ReLU\\
    2nd Conv2D (strided)&(328,400,16)& ReLU \\
    3rd Conv2D&(164,200,8)& ReLU \\
    4th Conv2D (strided)&(164,200,8)& ReLU \\
    5th Conv2D&(82,100,8)& ReLU \\
    6th Conv2D (strided)&(82,100,8)& ReLU \\
    1st ConvLSTM2D&(41,50,8)& ReLU \\
    2nd ConvLSTM2D&(41,50,8)& ReLU \\
    7th Conv2D&(41,50,8)& ReLU \\
    8th Conv2DTranspose (inverse)&(41,50,8)& ReLU \\
    9th Conv2D&(82,100,8)& ReLU \\
    10th Conv2DTranspose (inverse)&(82,100,8)& ReLU \\
    11th Conv2D&(164,200,8)& ReLU \\
    12th Conv2DTranspose (inverse)&(164,200,8)& ReLU \\
    13th Conv2D&(328,400,16)& ReLU \\
    Output/14th Conv2D&(328,400,4)& Linear \\
    \hline \hline
\end{tabular}
  \label{Case2Arch}
\end{center}
\end{table}

\chapter{Time Evolution of Additional Predicted fields}

   \begin{figure}[htbp!]
       \centering
        \includegraphics[width=0.98\linewidth, height=7cm]{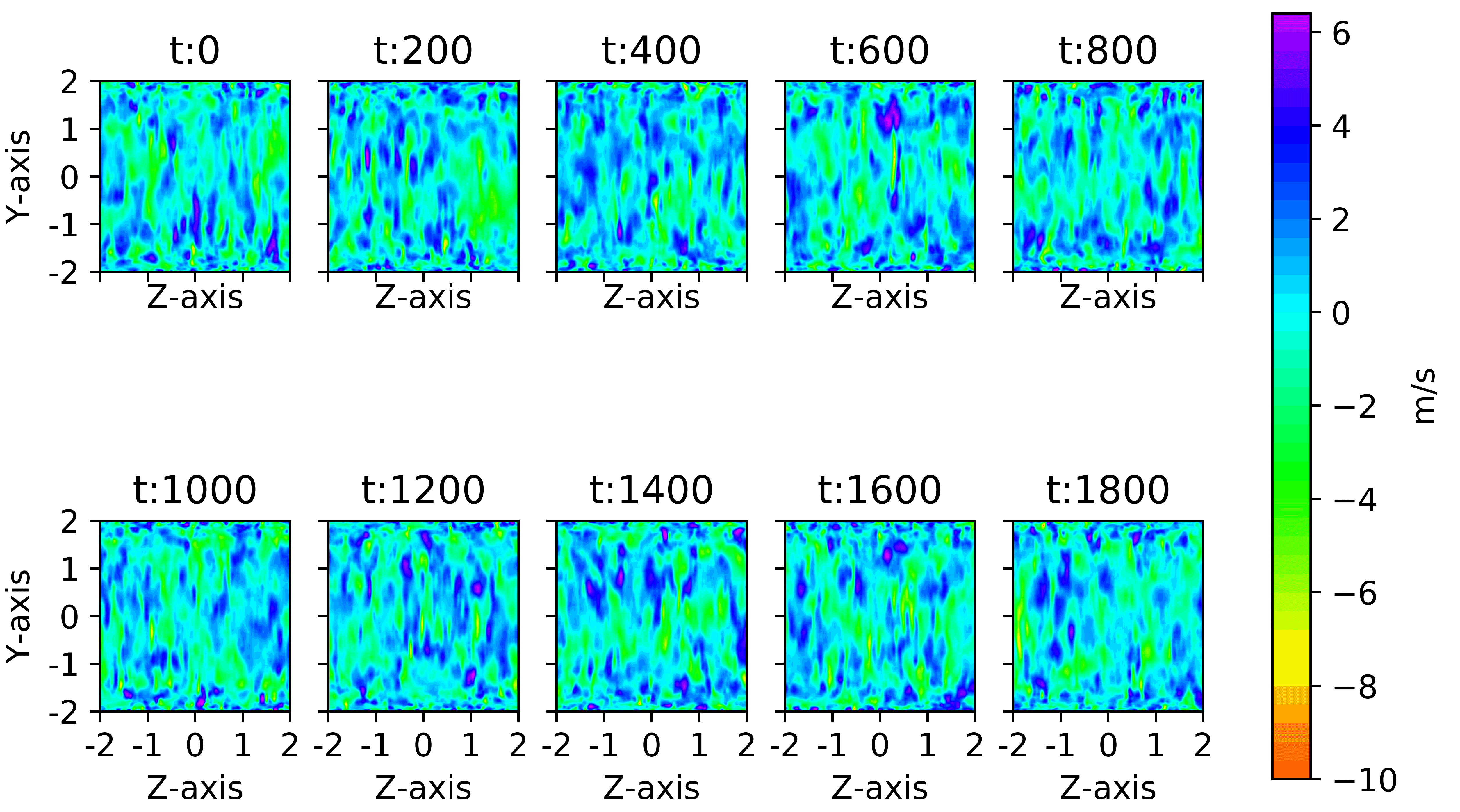} 
        \caption{Learned $u^\prime$ fluctuations for Case-1}
        \label{c1_plot_predicted_aPriori_ch0}
   \end{figure} 

   \begin{figure}[htbp!]
       \centering
        \includegraphics[width=0.98\linewidth, height=9cm]{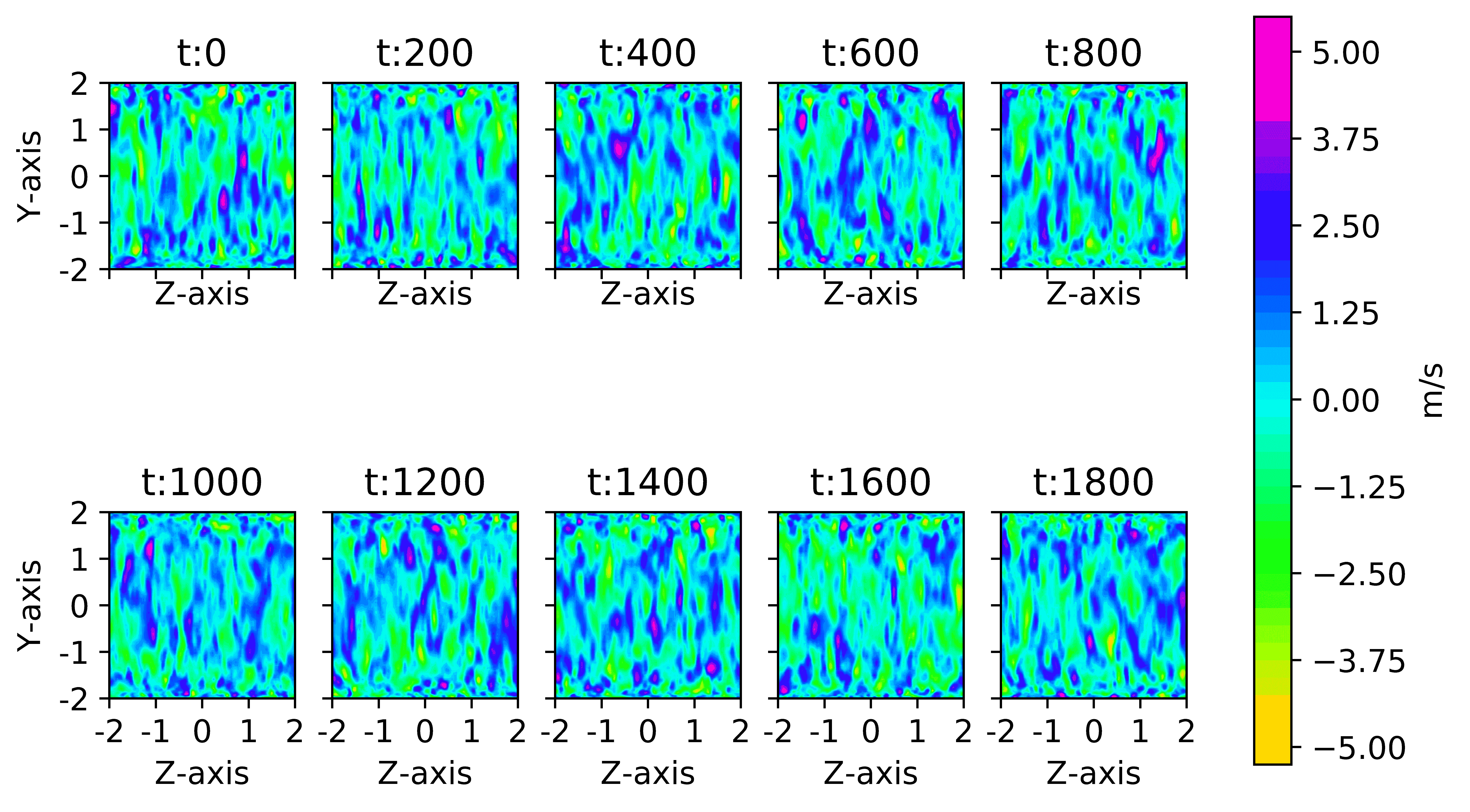}  
        \caption{Learned $v^\prime$ fluctuations for Case-1}
        \label{c1_plot_predicted_aPriori_ch1}
   \end{figure} 

   \begin{figure}[htbp!]
       \centering
        \includegraphics[width=0.98\linewidth, height=10cm]{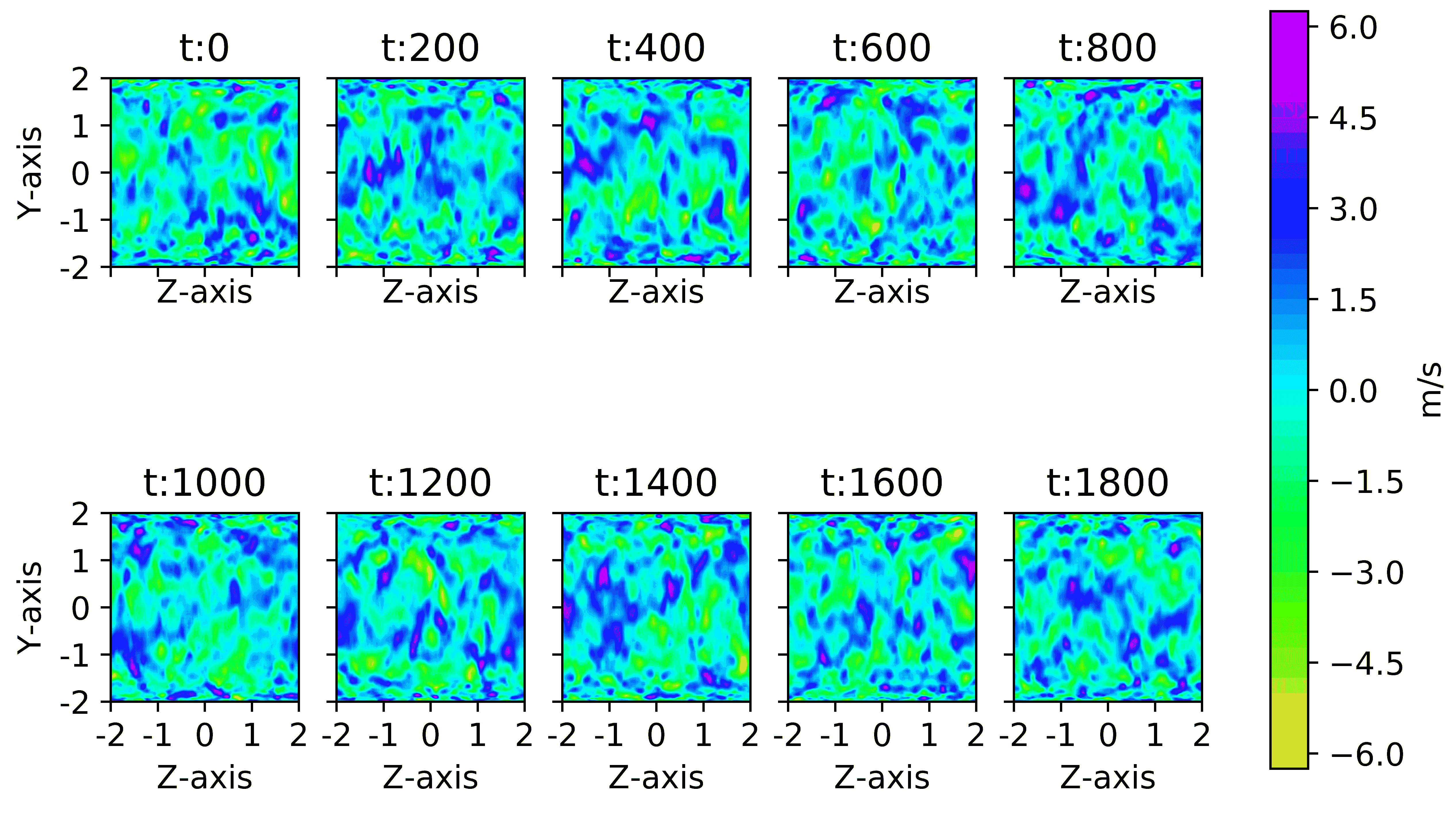} 
        \caption{Learned $w^\prime$ fluctuations for Case-1}
        \label{c1_plot_predicted_aPriori_ch2}
   \end{figure}

   \begin{figure}[htbp!]
       \centering
        \includegraphics[width=0.98\linewidth, height=10cm]{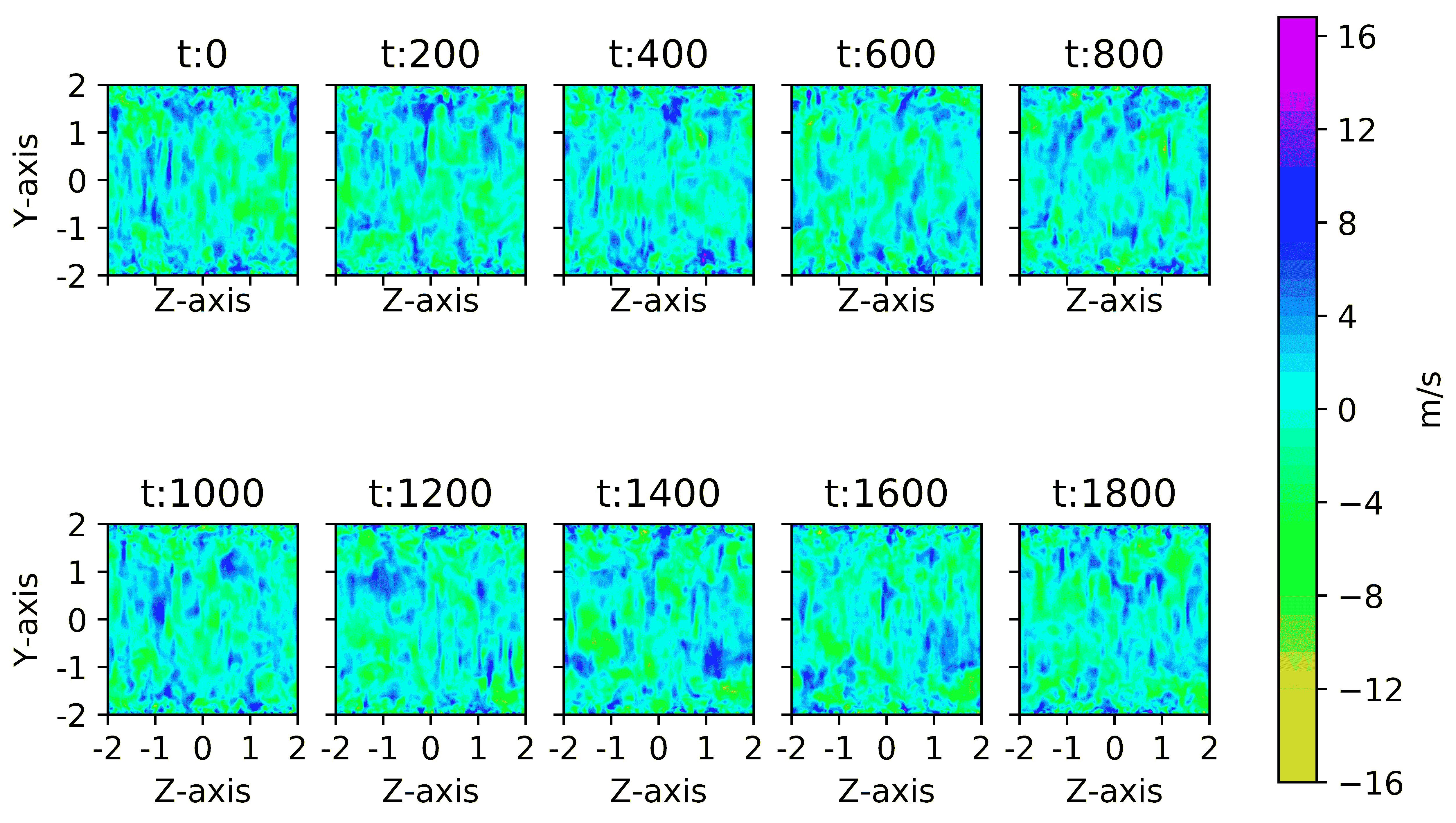} 
        \caption{Learned $u^\prime$ fluctuations for Case-2}
        \label{c2_plot_predicted_aPriori_ch0}
   \end{figure} 

   \begin{figure}[htbp!]
       \centering
        \includegraphics[width=0.98\linewidth, height=10cm]{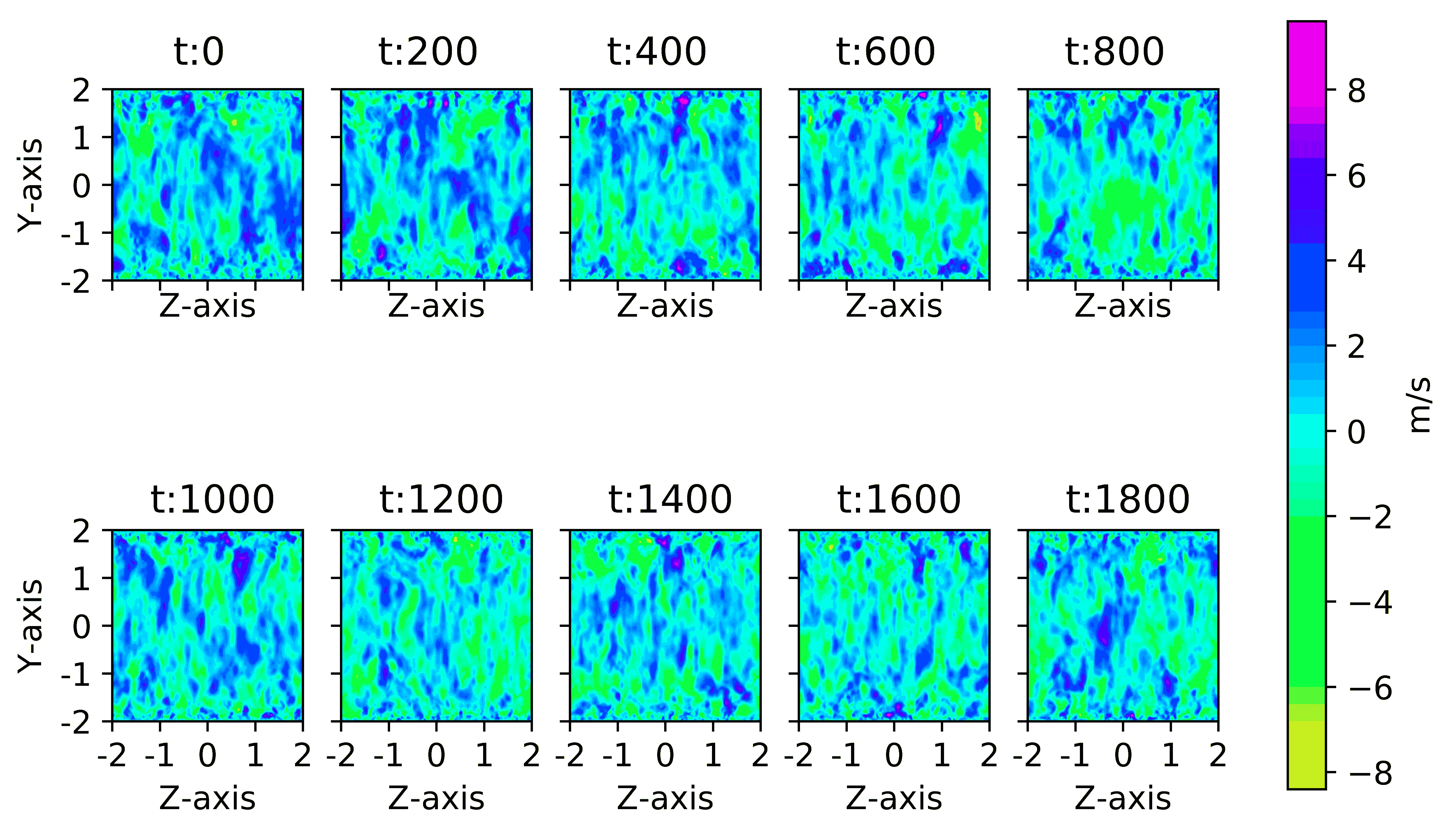} 
        \caption{Learned $v^\prime$ fluctuations for Case-2}
        \label{c2_plot_predicted_aPriori_ch1}
   \end{figure} 

   \begin{figure}[htbp!]
       \centering
        \includegraphics[width=0.98\linewidth, height=10cm]{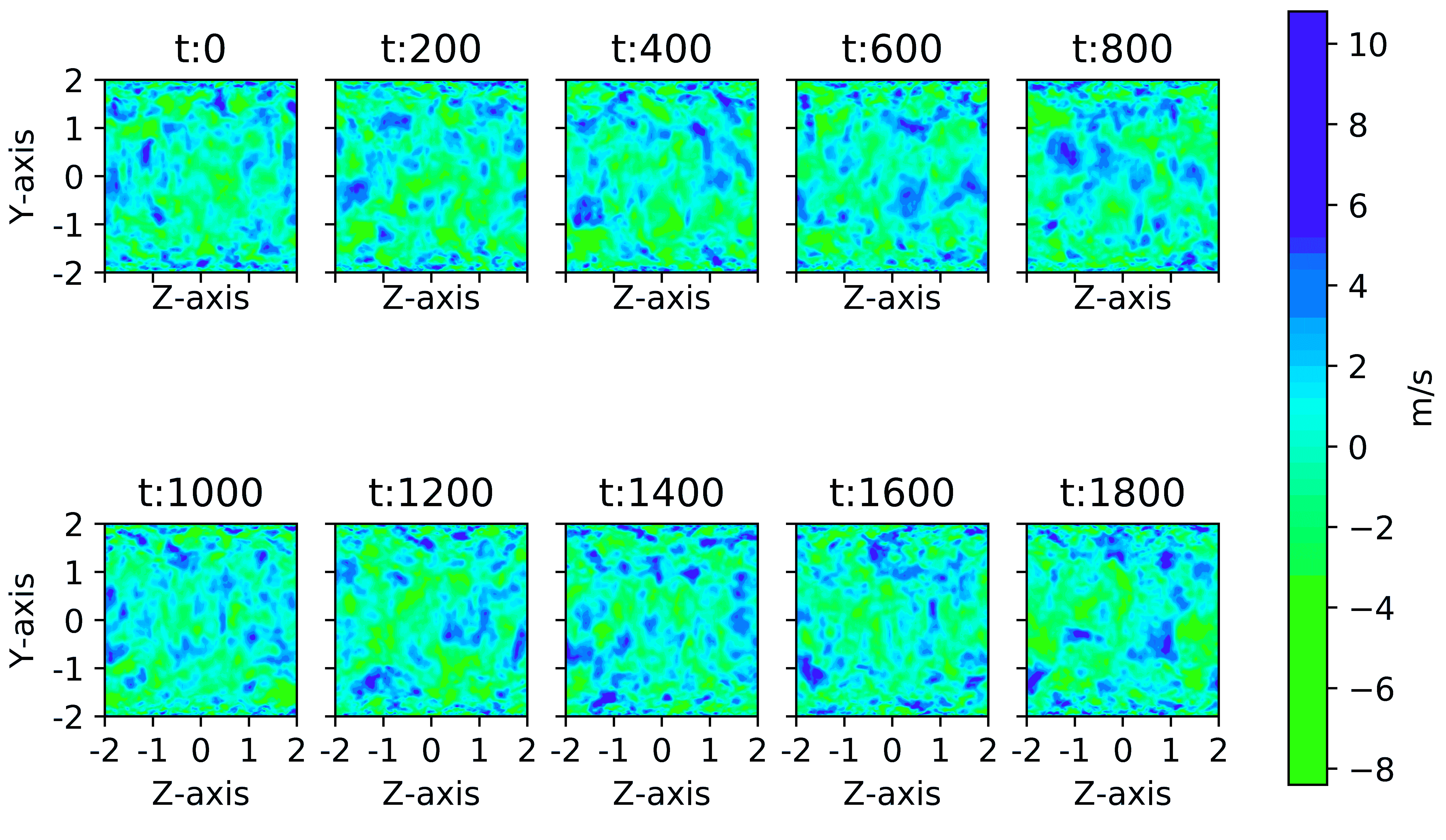} 
        \caption{Learned $w^\prime$ fluctuations for Case-2}
        \label{c2_plot_predicted_aPriori_ch2}
   \end{figure}

\end{document}